\documentclass[12pt,reqno]{article}
\usepackage[pdftex,colorlinks=true]{hyperref}
\hypersetup{citecolor=darkblue,linkcolor=darkblue,urlcolor=darkblue}
\usepackage[left=0.7in,top=1in,right=0.7in,bottom=1in,nohead]{geometry}
\usepackage[round]{natbib}
\usepackage{threeparttable,booktabs,ulem}
\usepackage[font={footnotesize,it}]{caption}
\usepackage{lscape,amssymb,amsmath,verbatim,graphicx}
\usepackage{epsfig,subfigure,float,subfigure,xcolor}
\usepackage{graphicx}
\usepackage{epsfig, colortbl,paralist}
\usepackage{multirow, microtype}
\usepackage{adjustbox,orcidlink}
\usepackage{lscape,lipsum}
\usepackage{xcolor}
\definecolor{darkblue}{rgb}{0,0,.6}
\definecolor{Gray}{gray}{0.9}

\newsavebox\CBox

\graphicspath{{plots/}}

\newcommand{\appendixnumberline}[1]{Appendix\space}

\let\oldappendix\appendix
\makeatletter
\renewcommand{\appendix}{%
  \addtocontents{toc}{\let\protect\numberline\protect\appendixnumberline}%
  \renewcommand{\@seccntformat}[1]{Appendix~\csname the##1\endcsname\quad}%
  \oldappendix
}
\makeatother

\DeclareMathOperator*{\argmin}{\arg\!\min}

\numberwithin{figure}{section}

\usepackage{algorithm, algorithmic}
\newcommand{\Rlogo}{\protect\includegraphics[height=1.8ex,keepaspectratio]{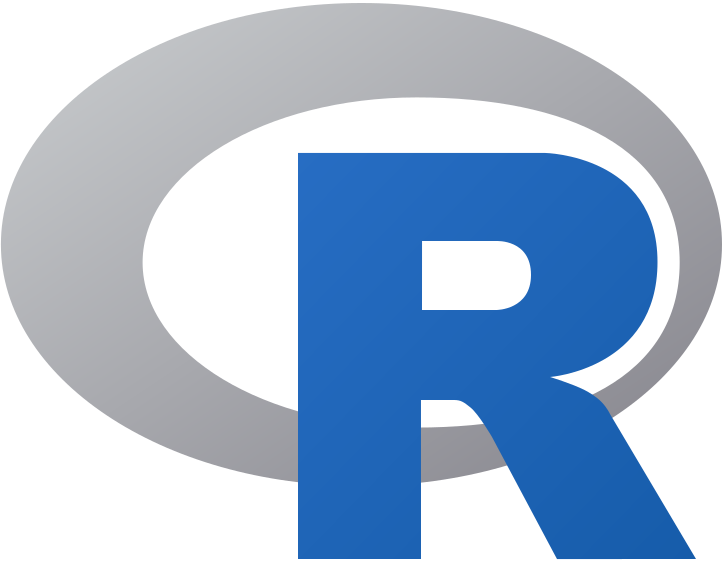}}

\graphicspath{{plots/}}

\numberwithin{equation}{section}
\numberwithin{theorem}{section}
\numberwithin{table}{section}

\usepackage{setspace}
\usepackage{newtxtext}
\usepackage{dsfont}
\newcommand{\blind}{0}

\usepackage{endnotes}
\onecolumn

\def\AmSTeX{$\cal A$\kern-.1667em\lower.5ex\hbox{$\cal M$}\kern-.125em
	$\cal S$-\TeX}
\def\BibTeX{{\rm B\kern-.05em{\sc i\kern-.025em b}\kern-.08em
		T\kern-.1667em\lower.7ex\hbox{E}\kern-.125emX}}
		
\begin{document}

\if0\blind
{
\title{Intraday FX Volatility-Curve Forecasting with Functional GARCH Approaches}
\author{
Fearghal Kearney \orcidlink{0000-0002-3251-8707}
\thanks{Finance and AI Lab, Queen's Business School, Queen's University Belfast, UK. BT9 5EE.}\\
Han Lin Shang \orcidlink{0000-0003-1769-6430}
\thanks{Department of Actuarial Studies and Business Analytics,
Macquarie University, Sydney, NSW 2109, Australia.}\\ 
Yuqian Zhao \orcidlink{0000-0002-5396-3316}
\thanks{Corresponding author. University of Sussex, Brighton, UK. BN1 9SN. Email: yuqian.zhao@sussex.ac.uk}
}
\date{}
\maketitle
} \fi

\if1\blind
{
   \title{Intraday FX Volatility-Curve Forecasting with Functional GARCH Approaches}
   \author{}
   \date{}
   \maketitle
} \fi

\begin{abstract}
This paper seeks to forecast intraday volatility curves for major foreign exchange (FX) currencies using functional GARCH models. Intraday return curves are observed at a daily frequency, yet preserve the full high-frequency trading structure, enabling volatility analysis at the intraday level. We demonstrate that the USD/EUR, USD/GBP, and USD/JPY intraday return curves exhibit strong cross-dependence, while individually they are serially uncorrelated but display long-range conditional heteroskedasticity. Embedding cross-currency dependence via multi-level functional principal component analysis and adding intraday bid-ask spread curves as exogenous drivers significantly improves intraday and day-ahead volatility forecasts relative to functional and realised-volatility baselines. The precise volatility forecasts motivate the construction of intraday Value-at-Risk (VaR). An intraday risk management application highlights that predicted intraday VaR curves can help mitigate dramatic losses in intraday trading strategies, showcasing their practical economic benefits in FX markets.
\medskip
\\
\noindent \textbf{Keywords:} Volatility modelling and forecasting, intraday return curves, intraday VaR, functional GARCH model, Forex market.\\
\textbf{JEL Classification:} C13, C32, G10
\end{abstract}

\newpage
\paragraph{Highlights}
\begin{itemize}
\item Forecast intraday FX volatility curves via functional GARCH-type models
\item Intraday FX curves display strong cross-dependence and long-range conditional heteroskedasticity
\item Modelling cross-currency dependence significantly improves forecast accuracy
\item Intraday bid-ask spread curves add predictive power
\item Intraday VaR curves provide a simple intraday risk management tool 
\end{itemize}

\newpage
\section{Introduction}\label{sec-intro}

The modelling and forecasting of foreign exchange (FX) market volatility has become increasingly important due to the growing volume of FX trading and recent volatility spikes, with implied volatility doubling (GBP), tripling (EUR), and quadrupling (JPY) for three reserve currencies previously renowned for their price stability in the post-COVID period.\footnote{BIS Triennial Central Bank Survey of Foreign Exchange and Over-The-Counter Derivatives Markets (Bank for International Settlements), \url{https://www.bis.org/publ/qtrpdf/r_qt2212f.pdf}; Chatham Financial Insights, \url{www.chathamfinancial.com/insights/market-volatility-impacts-fx-markets).}}
Downside risks and volatility spillovers across currencies present significant challenges for currency trading, particularly as activity shifts toward shorter intraday horizons \citep{chulia2018currency, rubaszek2025intraday}. Early studies document the growing demand for intraday risk management tools for high-frequency trading \citep[see, e.g.,][]{Engle00, AK08, AL10}. However, for decades, this demand remained largely unmet, as the field lacked suitable econometric tools and the literature primarily focused on inter-daily risk management. The primary challenge lies in forecasting FX volatility on the intraday horizon. This paper aims to enhance intraday risk management practices by proposing a methodological framework for forecasting the intraday volatility of three major currency pairs: USD/EUR, USD/GBP, and USD/JPY.
 
While forecasting FX volatility at the inter-daily horizon has been extensively studied \citep[see, e.g.,][]{martens2001forecasting, GM15, BKV16}, research on intraday volatility forecasting remains limited, as traditional models struggle to capture dynamics at such high frequencies. Early work by \cite{AB97} highlights the need for new approaches in this setting, concluding that standard GARCH-type models are inadequate for modelling intraday dynamics. Subsequently, realised volatility-based models have been developed to incorporate intraday trading information \citep[see, for example,][]{BS02, Sevi14, BKV16}. Nonetheless, these models still primarily forecast volatility at the inter-daily level by aggregating intraday information, rather than modelling intraday volatility directly.

To tackle this challenge, we consider FX intraday return curves, deemed functional data objects, observed at an inter-daily frequency with the entire intraday movements well preserved. The conditional volatility of intraday return curves can be modelled and forecast using recently developed functional GARCH-type (FGARCH hereafter) models. Intuitively, the conditional volatility inherits the functional form of the return curves, such that volatility itself is represented as an intraday curve, preserving fluctuations across the entire intraday trading horizon. The adoption of FGARCH-type models is motivated by their strong foundations in functional data analysis.\footnote{For monographs on functional data analysis theory and applications, we refer interested readers to \cite{Bosq00}, \cite{RS06}, \cite{HK12}, and \cite{KR21}.} Variants of FGARCH-type models have been developed over the past decade to capture volatility clustering dynamics in financial markets \citep[][]{HHR13, AHP17, CFH+19, RWZ20, RWZ21}.\footnote{\citeauthor{HHR13}'s \citeyearpar{HHR13} seminal work proposes a functional autoregressive conditional heteroscedasticity (ARCH) model, which is subsequently generalised to the FGARCH(1,1) model in \cite{AHP17} and the FGARCH$(p,q)$ model in \cite{CFH+19}. \cite{RWZ21} develops an FGARCH-X model to incorporate exogenous information with an additional X covariate.}

As intraday dynamics often exhibit complex comovements across global markets, direct applications of FGARCH models to the FX market may overlook important characteristics of FX volatility \citep{GM15, rubaszek2025intraday}. We find that the intraday return curves of USD/EUR, USD/GBP, and USD/JPY, exhibit strong cross dependence; however, when viewed as isolated time series, they are serially uncorrelated and display long-range conditional heteroskedasticity. Existing FGARCH-type models can only capture short-range dependence and explain the dynamics of the intraday volatility process in the univariate scenario, omitting their rich dependence structures, such as the long-range conditional heteroskedasticity and cross-asset dependence in FX markets.

Our paper attempts to address the aforementioned methodological issues comprehensively. We develop a parsimonious FGARCH model that incorporates cross-currency connectedness by employing multi-level functional principal component analysis (MFPCA) for dimension reduction. This approach avoids the need to extend heavily parameterized multivariate GARCH models into the functional time series framework. Instead, we retain the existing FGARCH-type specifications but estimate the models through a dimension-reduction procedure based on basis functions derived from MFPCA across multiple FX assets, thereby incorporating cross-dependence. Under the same estimation framework, we also introduce a long-range dependent functional principal component analysis (LFPCA) to account for the long-range conditional heteroskedasticity in FX assets. Both methods demonstrate superior performance in forecasting FX volatility.

Empirically, we forecast the intraday volatility of return curves, specifically the overnight cumulative intraday return (OCIDR) curves for three major FX pairs: USD/EUR, USD/GBP, and USD/JPY. This return object is particularly interesting to study because 
\begin{inparaenum}
\item[(1)] its smoothing feature makes it more suitable for functional time series modelling \citep{RWZ20b}, 
\item[(2)] it accounts for the overnight effect, which is crucial for volatility modelling \citep{HL06}, and 
\item[(3)] it provides a convenient framework for modelling the volatility of inter-daily closing returns at each intraday grid point.
\end{inparaenum}
To model the conditional volatility of the OCIDR curves, we consider the FGARCH(1,1) model and the functional GARCH-X (FGARCH-X) model \citep{RWZ21}, which incorporates intraday bid-ask spread microstructure information. Our findings show that taking into account the connectedness between multiple currencies sharpens the predictability of the conditional volatility of intraday return curves. Additionally, incorporating intraday bid-ask spread information further improves FX volatility forecasting. 

Building on precise intraday volatility forecasts, we introduce an intraday risk management framework by constructing intraday value-at-risk (VaR) curves. This approach mitigates sharp losses by reducing the maximum drawdown of a benchmark intraday trading strategy during extreme market events. The gains are strongest in high-uncertainty states, consistent with recent evidence on intraday connectedness and bad-shock asymmetries in FX markets.

Our four main contributions are as follows: 
\begin{inparaenum}
\item[1)] We observe from the empirical data that the FX intraday returns are highly cross-dependent and exhibit long-range conditional heteroskedasticity. To accommodate these features in the existing FGARCH-type models, we propose two new data-driven basis methods for the dimension-reduction process in estimating the FGARCH-type models, namely, multi-level FPCA and long-range FPCA methods, which preserve the properties of cross-dependence of the return curves and their long-range dependence, respectively. Both methods generally show improved out-of-sample performance compared to the standard static and dynamic FPCA dimensional reduction methods commonly used in the literature. This is particularly evident in the multi-level FPCA approach, which uses common and currency-specific factors in FX currencies.  
\item[2)] We model and forecast the conditional volatility of the intraday FX return curves through an FGARCH-X model. The FGARCH-X model, which incorporates exogenous microstructure information -- intraday bid-ask spreads, enhances volatility forecasting by improving the explainability of long-range conditional heteroskedasticity and adjusting the level of volatility forecasts. Our empirical results suggest that, consistent with inter-daily volatility forecasting, the bid-ask spread information enhances the predictability of the intraday returns' conditional volatility. 
\item[3)] We statistically compare the relative performance of the volatility forecasting model candidates using the robust loss function approach of \cite{P2011} combined with \citeauthor{DM1995}'s \citeyearpar{DM1995} test and the model confidence set of \cite{HLN11}. This comparison was complicated by latent intraday volatility and the lack of a consistent estimator.
\item[4)] To explore the risk management implications of intraday volatility forecasting and to economically evaluate our out-of-sample forecasts, we compute intraday VaR using predicted values of conditional volatility. We uncover improvements in annual returns, Sharpe ratios, and maximum drawdowns using a VaR-informed intraday trading strategy. The trading strategy successfully avoids dramatic declines in intraday trading activities during extreme market conditions.
\end{inparaenum} 

The paper is structured as follows. We begin by describing our dataset and discussing the properties of FX intraday returns and bid-ask spreads curves in Section~\ref{sec-data}. Section~\ref{sec-method} outlines existing FGARCH-type models and introduces new data-driven bases in dimension reduction for adapting FX intraday volatility forecasting. Section~\ref{sec-app} presents the statistical forecast performance of the models, and Section~\ref{sec-port} provides intraday risk management applications. We conclude and make a recommendation for future work in Section~\ref{sec-con}.

\section{FX data and preliminary analysis}\label{sec-data}

This section introduces the dataset and presents preliminary analyses of its properties. We consider the three most traded currencies against USD, namely EUR, GBP, and JPY. These currencies constituted over 75\% of the turnover of OTC FX instruments in 2019, according to the Bank for International Settlements.\footnote{Bank of International Settlements, \url{https://stats.bis.org/statx/srs/table/d11.3}} For each currency, the intraday spot, bid, and ask rates are collected at a 5-minute frequency from a collection of databases, including Morningstar, FXCM, TenFore, and Thomson Reuters. The sample ranges from 07-January-2014 to 30-September-2020.\footnote{Due to limited availability of high-frequency bid-ask spread data, the analysis is not extended to more recent periods, where FGARCH-X models with more recent bid-ask spread covariates would be required. In unreported results using a more recent sample, FGARCH models that only require intraday returns yield findings consistent with those reported here. The chosen sample, therefore, aligns to ensure comparability between the FGARCH and FGARCH-X specifications.} Given the OTC nature of FX markets, our sample covers U.S. trading hours from midnight on Monday until the close on Thursday night. Intraday data on Sunday and Friday are truncated from our sample, as they do not form 24-hour trading sessions. Consequently, each FX rate forms an information set of 1,275 trading days, each with 288 intraday grid points. In an unreported analysis, we also include forecasting during the U.S. main trading session from 7:00 to 17:00, Monday to Friday, spanning 1,722 trading days with 108 daily grid points each day. The results show consistency with the dataset used in this study.

To study the volatility of intraday return curves, we calculate the overnight cumulative intraday return curves (OCIDR), $y_t(u)$, as:
\begin{equation}\label{eq-ocidr}
y_t(u) = [\ln P_t(u)- \ln P_{t-1}(1)] \times 100,
\end{equation}
where $P_t(u)$ denotes the spot rate of either currency on day $t$ evaluated at intraday time $u$, $u\in[0,1]$, and $P_{t-1}(1)$ is the closing spot rate on day $t-1$. The functional data objects can be obtained using smoothing techniques \citep[][Chapter~3]{RS06} or simple linear interpolation under the condition of continuity \citep{RWZ20b}. The OCIDR curves provide a continuous trajectory, so they are suitable for casting as functional time series objects while accounting for the overnight effect \citep{HL06}. Desirably, forecasting the volatility of OCIDR curves includes inter-daily volatility prediction because OCIDR curves end at daily closing returns, defined as $y_t = \ln P_t(1)- \ln P_{t-1}(1)$.

Additionally, we obtain intraday bid-ask spread curves. The inclusion of the bid-ask spread is motivated by \cite{ER19}, who point out that although the FX rate is commonly represented as the midpoint between the best bid and the ask price, in reality, all FX trades are executed at the bid or the ask price. This bid or ask trading dynamic also brings about imperfect measurement in determining the variance of return, further motivating our functional characterisation of the intraday data. The use of bid-ask spreads has been widely discussed in the volatility forecasting literature. For example, \citet{BM94} state that bid-ask spreads act as a measure of liquidity and have a significant relationship with the volatility of the FX rate. Incorporating bid-ask spread data also helps address the issue of bid-ask bounce, which might be misinterpreted as volatility arising from price movements. As a measure of liquidity, we calculate the intraday bid-ask spread curves (IBAS) through
\begin{equation*}
\text{IBAS}_t(u) = \text{ask}_t(u) - \text{bid}_t(u), \quad u\in[0,1].
\end{equation*}
To achieve stationarity and generate an information flow that is compatible with the OCIDR curve, we calculate the overnight cumulative IBAS (OCIBAS) curve using $\text{OCIBAS}_t(u) = \text{IBAS}_t(u)- \text{IBAS}_{t-1}(1)$. The OCIBAS curve is also assumed to be square integrable with the sample path in $\mathcal{L}^2[0,1]$. Intuitively, higher values of the $\text{OCIBAS}_t(u)$ curve represent lower liquidity and vice versa.
\begin{figure}[H]
\centering
\caption{Plots of Overnight Cumulative Intraday Return (OCIDR) curves and Cumulative Intraday Bid-Ask Spread (OCIBAS) curves for USD/EUR. The x-axis represents each 5-minute frequency intraday grid point, with the y-axis representing returns and spread, respectively, for the 1,275 trading days in our 07-January-2014 to 30-September-2020 sample. OCIDR and OCIBAS are defined in Section~\ref{sec-data}, with higher OCIBAS values representing lower levels of liquidity and vice versa.} 
\label{figure-ocidr}
\includegraphics[width=8.45cm]{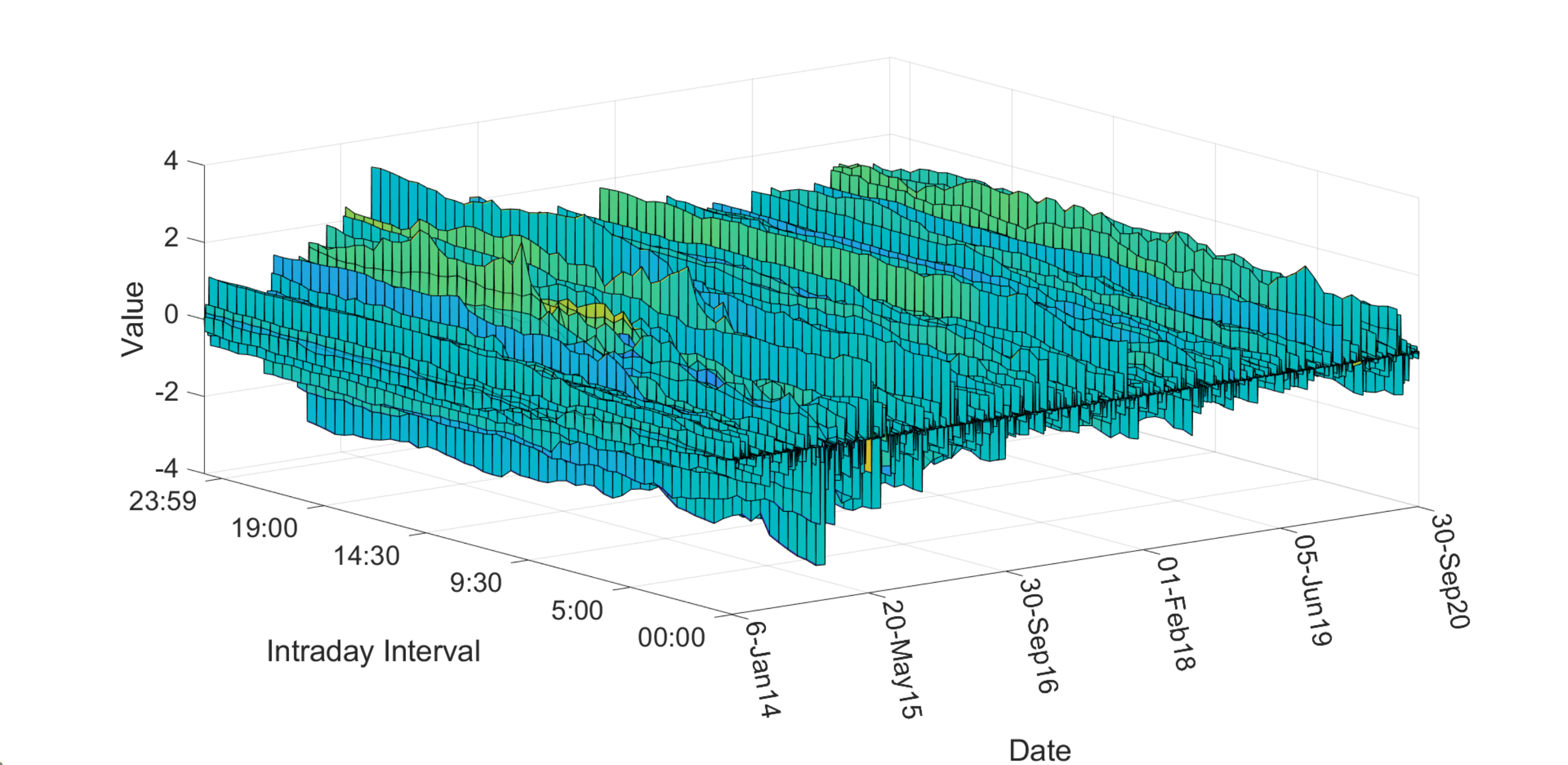}
\quad
\includegraphics[width=8.45cm]{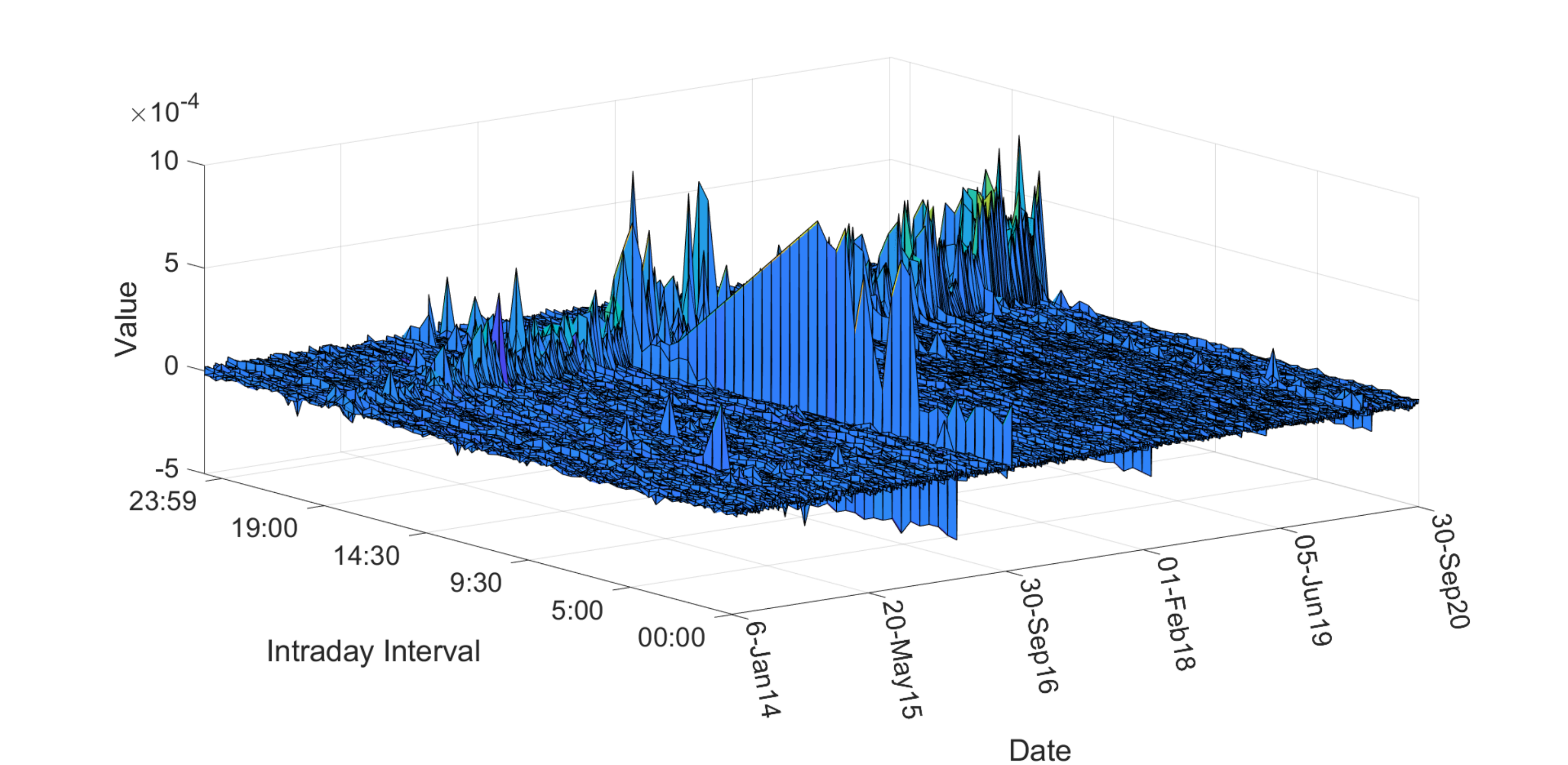}
\end{figure}

Figure~\ref{figure-ocidr} displays the plots of the OCIDR and OCIBAS curves for USD/EUR. Each OCIDR curve records the evolution of the return movements over the trading day. The plot of the OCIBAS curves shows that they spike during opening and closing trading hours, implying less liquidity during these trading sessions. To understand the properties of these curve processes, we deploy the KPSS-type stationary test \citep{HKR14}, the static FPCA-type normality test \citep{GHH+18}, the independence test \citep{KRS17}, and the conditional heteroskedasticity test \citep{RWZ20}. Table~\ref{table-data} documents the \textit{p}-values for each statistical test in the first training sample, which runs from 03-January-2014 to 31-December-2016. We limit our testing to the first training sample for in-sample analysis throughout the paper, and to ensure the robustness of our reported results, we usually replicate the analysis for other training samples. Table~\ref{table-data} indicates that the OCIDR curves are weakly stationary, non-Gaussian, serially uncorrelated, and conditionally heteroscedastic functional time series. Interestingly, the properties of the OCIDR curves uncovered here for the FX market are consistent with intraday return curves in the equity and commodity markets \citep[see, e.g.,][]{RWZ21}.
\begin{table}[H]
\centering
\caption{$P$-values of statistical tests on Overnight Cumulative Intraday Return (OCIDR) curves and Overight Cumulative Intraday Bid-Ask Spread (OCIBAS) curves, as defined in Section~\ref{sec-data}, for our three currency pairs, USD/EUR, USD/GBP, and USD/JPY. The results are calculated for 07-January-2014 to 17-May-2017, the first training sample period of our data. We test for stationarity, normality, autocorrelation, and heteroskedasticity with the $\widehat{a}_x$ and $\widehat{a}_{x^2}$/$\widehat{a}_{|x|}$ representing the short- and long-range dependency memory parameter estimates obtained using the Local Whittle (LW) estimator. Lags of 1, 5, 10, and 20 days are specified for the autocorrelation and heteroskedasticity tests.} \label{table-data}
\begin{adjustbox}{max width=\linewidth}
\begin{tabular}{@{}lcccccccccccc@{}}
\toprule
		& Stationarity & Normality & \multicolumn{4}{c}{Autocorrelation} & \multicolumn{4}{c}{Heteroskedasticity} & $\widehat{a}_x$ & $\widehat{a}_{x^2}$ \\\midrule
		Lag           &            &           & 1       & 5       & 10     & 20     & 1        & 5       & 10      & 20      &           &                              \\\hline
		$\mbox{OCIDR}_{\text{USD/EUR}}$ &   0.22     & 0.00      &  0.71  &  0.61  &  0.43  &  0.60  & 0.04     & 0.00    & 0.00    & 0.00    &   0.02   &   0.20                       \\
		$\mbox{OCIDR}_{\text{USD/GBP}}$ &   0.60     & 0.00      &  0.64  &  0.68  &  0.56  &  0.40  & 0.06     & 0.00    & 0.00    & 0.00    &   -0.04   &   0.33                    \\
		$\mbox{OCIDR}_{\text{USD/JPY}}$ &   0.68     & 0.00      &  0.19  &  0.10  &  0.28  &  0.25  & 0.00     & 0.02    & 0.04    & 0.02    &   0.06   &   0.13                  \\&   & \multicolumn{4}{c}{ } & \multicolumn{5}{c}{ } &   & $\widehat{a}_{|x|}$  \\
		$\mbox{OCIBAS}_{\text{USD/EUR}}$ &   0.11     & 0.00      &  0.37  &  0.25   &  0.36  &   0.37   &   0.96   &  0.09   &   0.01  &   0.00  &     0.11 &     0.12                 \\
		$\mbox{OCIBAS}_{\text{USD/GBP}}$  &   0.04     & 0.00      &  0.14  &  0.37   &  0.26  &  0.20    &   0.95   &     0.07   &  0.00   &  0.00   &     0.18  &  0.19                  \\
		$\mbox{OCIBAS}_{\text{USD/JPY}}$  &   0.07     & 0.00      &  0.61  &  0.83   &  0.34  &   0.34   &   0.00   & 0.00    &   0.00  &  0.00   &  0.10    &        0.27             \\\bottomrule
\end{tabular}
\end{adjustbox}
\end{table}

We also investigate the long-range dependence structure of the level and squares of the FX curve sequences. According to \citet{LRS20, LRS21}, the memory parameter $a$ of the stationary curves can be estimated using either via a semiparametric R/S method or a feasible local Whittle (LW) estimator. Taking into account the efficiency of the estimator, we apply the latter method as suggested by \cite{LRS21}, with the LW estimator $\widehat{a}_x$ denoting the memory parameter of the level curves, and $\widehat{a}_{x^2}$ denoting the memory parameter of the squared process. Estimates of $\widehat{a}_x$ and $\widehat{a}_{x^2}$ show that the levels of the OCIDR curves are short-range dependent, while the squared processes are long-range dependent. The absolute processes $\widehat{a}_{|x|}$ of OCIBAS curves are also long-range dependent.

Furthermore, it has been documented that co-movements between excess returns of a variety of internationally traded financial assets are widespread \citep[see, e.g.][]{Bekaert95}. These studies motivate a series of works on forecasting daily volatility in the FX market. Recently, theoretical bases for understanding cross-currency co-movements have been built by \cite{GM15} and \cite{CCGR18}, with \cite{GM15} further highlighting the role of co-movements in modelling FX volatility. This cross-dependence structure between related currency pairs may also add predictability to return volatility in FX markets. Therefore, we seek to understand the interdependency between currency pairs by estimating the sample covariance operator between their OCIDR curves.

\begin{figure}[H]
\centering
\caption{Estimated covariance between pairwise Overnight Cumulative Intraday Return (OCIDR) curves obtained using their sample covariance operators. The various pairwise combinations of USD/EUR, USD/GBP, and USD/JPY are presented. The $x$-axis and $y$-axis show the intraday time points for each currency pair, with the $z$-axis indicating the correlation level between the intersection of those time points for that currency pair combination. The results are calculated using the whole sample period.}\label{figure-corr}
\includegraphics[width=8.5cm]{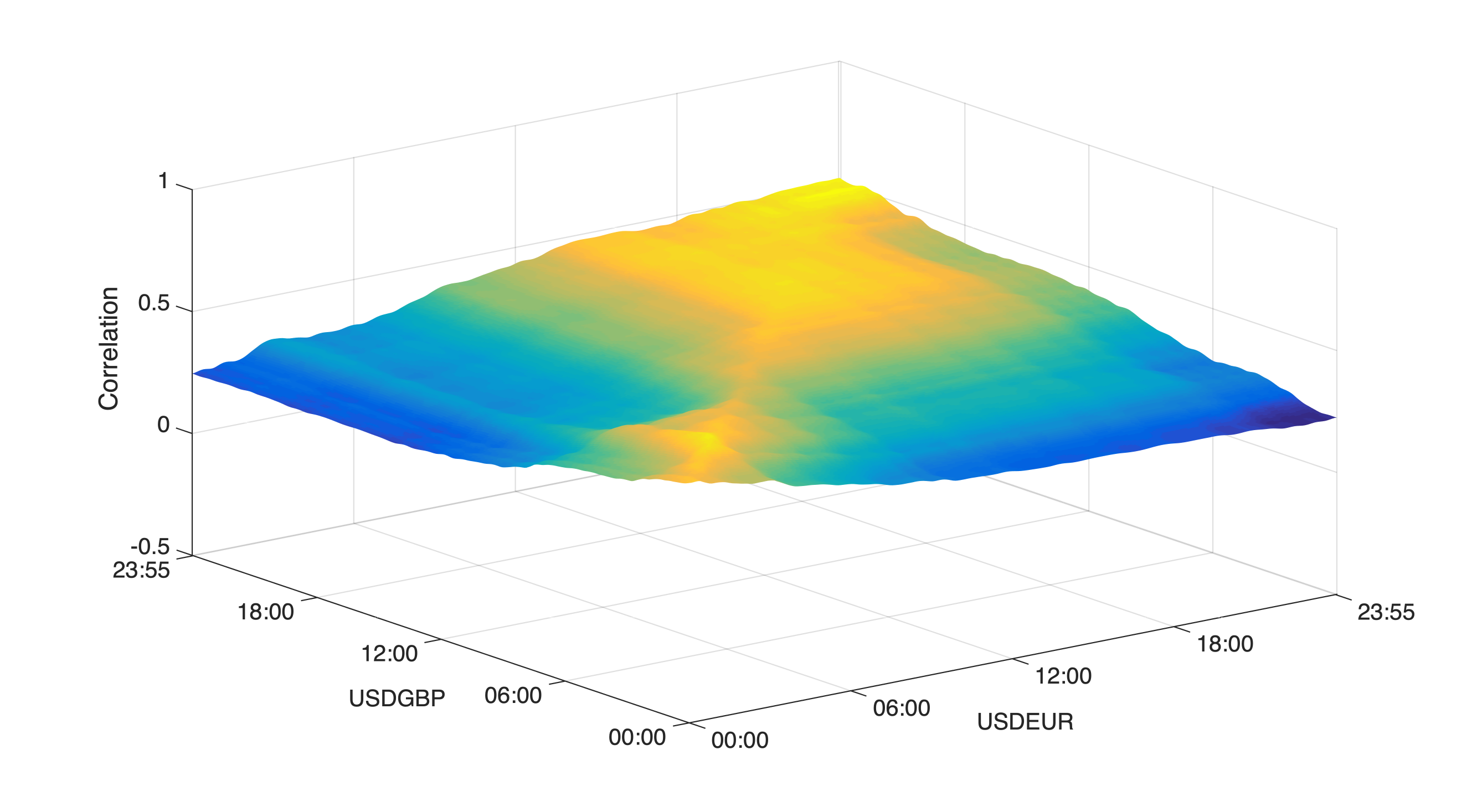}\quad
\includegraphics[width=8.5cm]{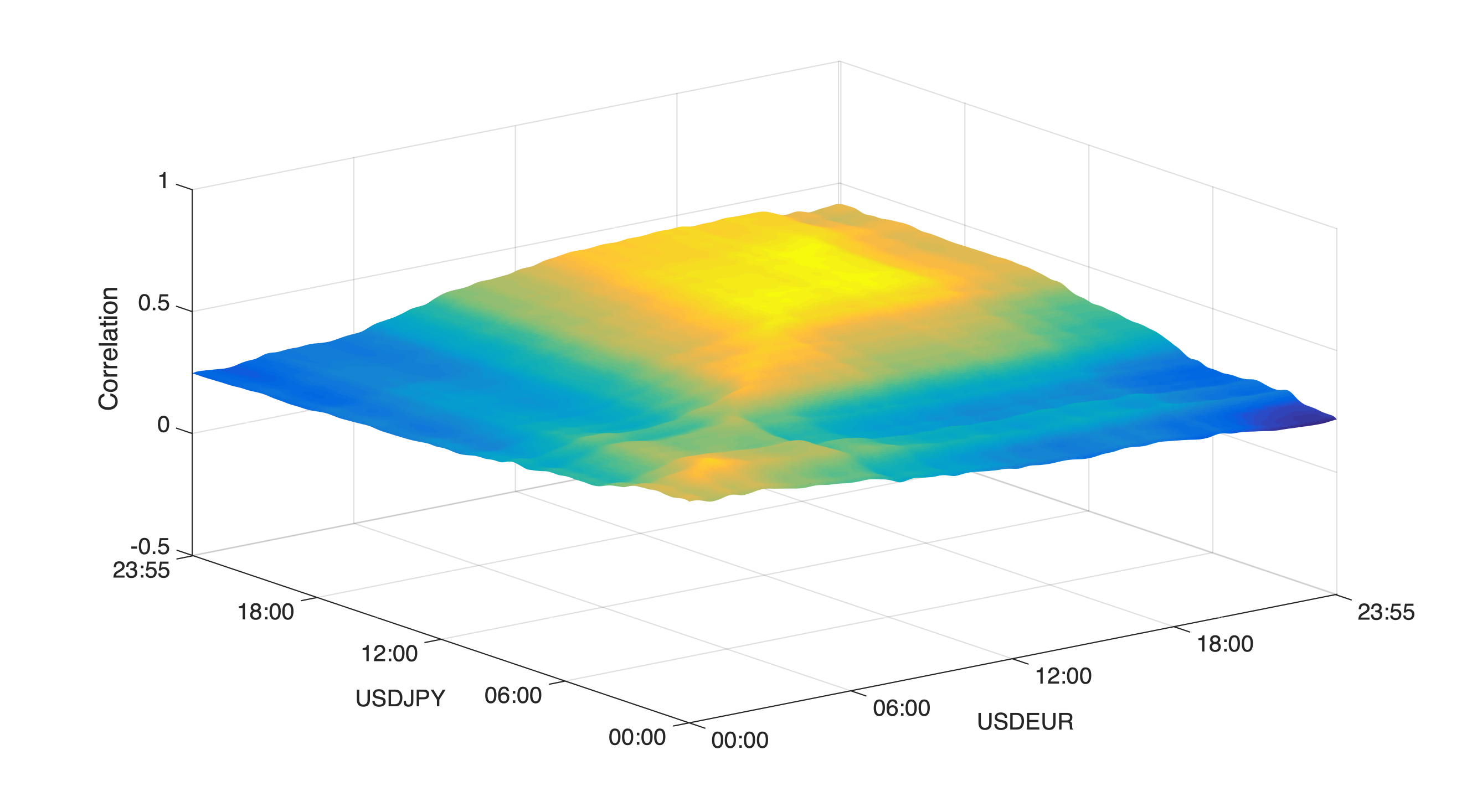}\quad
\includegraphics[width=8.5cm]{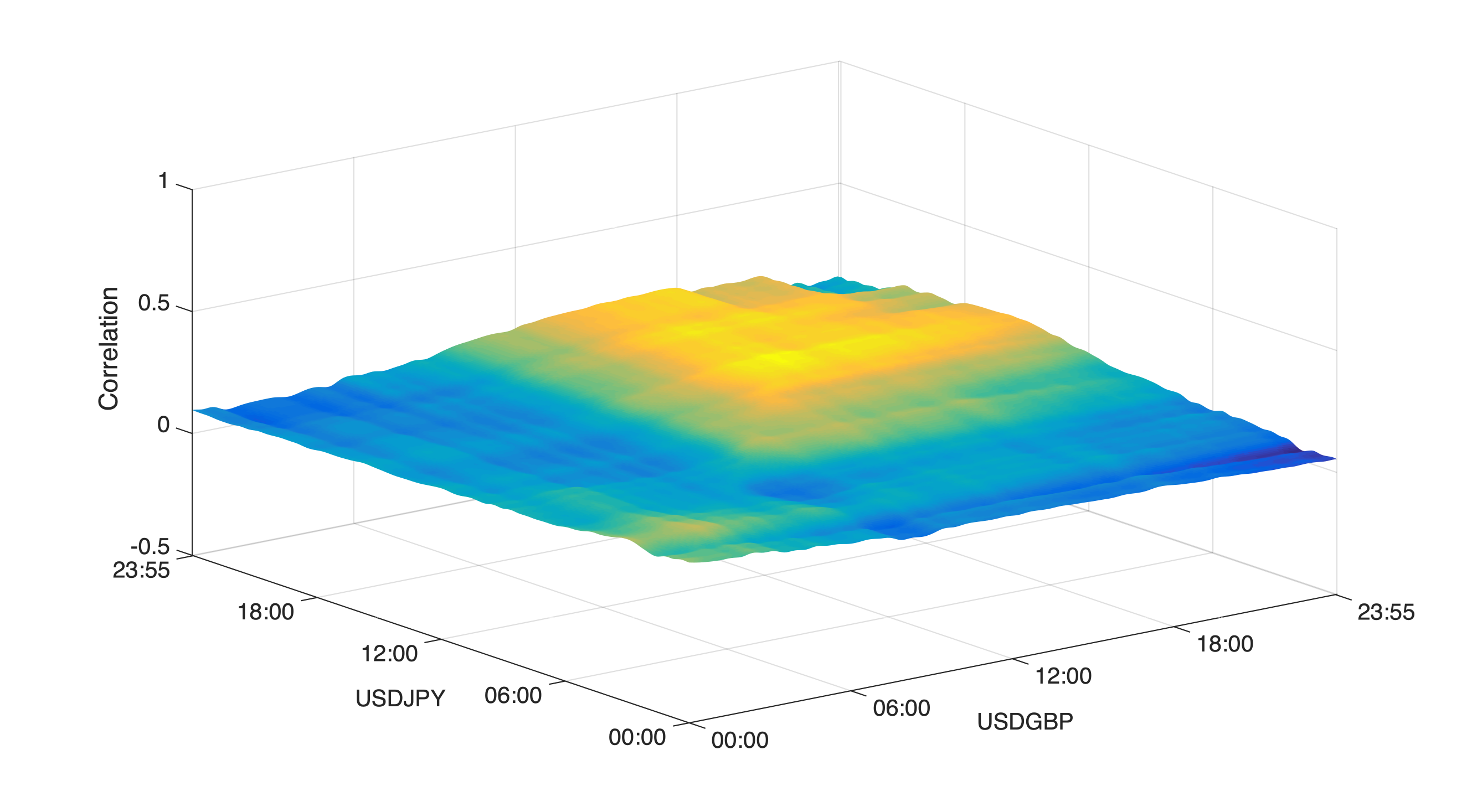}
\end{figure}

Figure~\ref{figure-corr} displays the estimated covariance operator between the various pairwise OCIDR curves. The results show that the three pairs of OCIDR curves are correlated, with a particularly high correlation between USD/EUR and USD/GBP, consistent with the dynamic connectedness of \cite{rubaszek2025intraday}. The source of this cross-dependency can be attributed to shocks to the U.S. economy and U.S. interest rates, which, given the base currency of USD, impact all FX rates we consider. \cite{EE95} provide empirical evidence that U.S. monetary policy shocks, such as interest rate changes, have significant spillover effects on exchange rates globally, highlighting how U.S. economic shifts create dependencies among various currencies. \cite{HW11} examine the impact of U.S. Federal Reserve announcements, particularly those concerning interest rates, on global asset prices, including exchange rates. This study supports the idea that cross-dependencies in currency markets often arise from U.S. economic shocks. Thus, the empirical finding of strong covariance motivates our position that accounting for \textit{cross dependence} may enhance volatility modelling and forecasting.

\section{Methodology}\label{sec-method}

\subsection{Functional GARCH-type models}\label{sec-model}

This section outlines the existing functional GARCH-type models that are used to fit and forecast the conditional heteroskedasticity of intraday return curves, denoted as $\{ y_t(u) \}$, where $u \in [0,1]$ represents the intraday time index, and $1 \leq t \leq N$ denotes the inter-daily time index. We assume that the intraday return curve, $y_t(u)$, is square-integrable and constitutes sample paths in a $\mathcal{L}^2[0,1]$ Hilbert space. In addition, it is equipped with an inner product $\left \langle y_1, y_2 \right \rangle=\int y_1(u)y_2(u)du$, resulting in the norm $||y(u)||=[\int y^2(u)du]^{1/2}$, for $\int \equiv \int_0^1$. For convenience, we also denote the integral operator of the kernel $\mathbf{g}(y)(u)=\int g(u,v) y(v) dv$, for $g(u,v) \in \mathcal{L}^2[0,1]^2$, and further define subspaces $\mathcal{L}^2[0,1]^+=\{y\in \mathcal{L}^2[0,1], y(u) \geq 0\}$ and $\mathcal{L}^2[0,1]_*^+=\{y\in \mathcal{L}^2[0,1], y(u)> 0\}$, for almost every~$u \in [0,1]$.

The intraday return curves can be expressed as $y_t(u)=c(u)+\widetilde{y}_t(u)$, where $c(u)$ represents the functional mean, and the mean-zero curve $\widetilde{y}_t(u)$ is the focus of interest. Given the increased prevalence of volatility clustering in data like $\widetilde{y}_t(u)$, \citet{HHR13} first propose and establish the stationary solution of the FARCH(1) model for the conditional volatility of $\widetilde{y}_t(u)$. However, this model has been shown to inadequately account for the conditional heteroskedasticity observed in financial assets \citep{RWZ20}. \citet{AHP17} and \citet{CFH+19} generalise the FARCH model and specify the conditional volatility of $\widetilde{y}_t(u)$ following the recursion of,
\begin{equation}\label{model-fgarch}
\begin{split}
& \widetilde{y}_t(u) =\sigma_t(u)\varepsilon_t(u), \quad u \in [0,1], \quad \mbox{ } 1 \leq t \leq N,\\
&\sigma^2_t(u)=\omega(u)+\sum_{j=1}^q\boldsymbol{\alpha}_j(\widetilde{y}^2_{t-j})(u)+\sum_{k=1}^p\boldsymbol{\beta}_k(\sigma^2_{t-k})(u),\\
& \qquad \mbox{ } = \omega(u)+\sum_{j=1}^q \int \alpha_j(u,v) \widetilde{y}_{t-j}^2(v) dv +
\sum_{k=1}^p \int \beta_k(u,v) \sigma_{t-k}^2(v) dv,
\end{split}
\end{equation}
where $\sigma^2_t(u)$ is a latent conditional volatility function. The error term $\epsilon_t(u)$ is assumed to be a stationary process with weak dependence with $\text{E}(\epsilon_t(u))=0$ and $\text{E}(\epsilon^2_t(u))=1$, and $\text{E}[\epsilon^2_0(u)]=1$ for all $\epsilon \in[0,1]$. It is needed for the identifiability condition. The coefficient function $\omega(u)\in \mathcal{L}^2[0,1]_*^+$, and the kernels of the coefficient operators, $\alpha_j(u,v)$ and $\beta_k(u,v)$ are elements in $\mathcal{L}^2[0,1]^+\times \mathcal{L}^2[0,1]^+$, for $1\leq j \leq q$ and $1\leq k \leq p$. We refer interested readers to \citet{CFH+19} for the stationary solution and the consistent estimation method of the FGARCH$(p,q)$ model. Appropriate lag orders $p$ and $q$ can be selected using a goodness-of-fit test for conditional heteroskedasticity \citep{RWZ20}, which shows that the FGARCH(1,1) model adequately captures the conditional heteroskedasticity of intraday return curves for many financial assets.

Another parsimonious framework that models conditional volatility dependence is the FGARCH-X model. \citet{RWZ21} extends the FGARCH(1,1) to an FGARCH-X model that uses an exogenous covariate X to add explanatory ability to the rich dynamic of conditional heteroskedasticity. The model specifies the conditional volatility $\sigma_t^2(u)$:
\begin{equation}\label{model-fgarchx}
\begin{split}
&\sigma_t^2(u)=\omega(u)+\boldsymbol{\alpha}(\widetilde{y}^2_{t-1})(u)+\boldsymbol{\beta}(\sigma^2_{t-1})(u)+\boldsymbol{\gamma}(x_{t-1})(u),\\
& \qquad \mbox{ } = \omega(u)+ \int \alpha(u,v) \widetilde{y}_{t-1}^2(v) dv +
\int \beta(u,v) \sigma_{t-1}^2(v) dv +
\int \gamma(u,v) x_{t-1}(v) dv,
\end{split}
\end{equation}
where $[\omega(u)+\boldsymbol{\gamma}(x_{t-1})(u)]\in \mathcal{L}^2[0,1]^+_*$ for all $t$, and the kernel of the coefficient operators $\alpha(t,s)$, $\beta(t,s)$ and $\gamma(t,s) \in \mathcal{L}^2[0,1]^+\times \mathcal{L}^2[0,1]^+$. The model has a stationary solution if the exogenous covariate is weakly stationary. According to \citet{RWZ21}, the long-range conditional heteroskedasticity may be accounted for by incorporating a long-range dependent covariate X. Models~\eqref{model-fgarch} and \eqref{model-fgarchx} serve as forecasting models once a dataset is fitted and the model coefficients are estimated.

To consistently estimate FGARCH-type models, two methods have been documented in the literature, namely ordinary least squares (OLS) \citep{AHP17} and quasi-maximum likelihood estimation (QMLE) \citep{CFH+19}. Given the advantages of efficiency estimation outlined in \citet{CFH+19}, we choose to estimate our models using the QMLE approach. Like all functional time-series models, the FGARCH-type models are intrinsically evaluated in an infinite-dimensional Hilbert space, meaning that they can only be estimated in a finite-dimensional space by way of dimension reduction. Following \citet{CFH+19}, we project the infinite-dimensional conditional volatilities $\sigma_t^2(u)$ into a finite $K$-dimensional subspace of $\mathcal{L}^2[0,1]^+$. Let $\{\psi_1, \psi_2, \dots, \psi_K\} \in \mathcal{L}^2[0,1]^+$, in which we denote $K$ as known linearly independent basis functions. We then project the kernel operators in models~\eqref{model-fgarch} and~\eqref{model-fgarchx} onto these bases. For example, for the model~\eqref{model-fgarchx} we have,
\begin{equation}\label{model-coefficient}
\omega=\sum_{l=1}^K d_l \psi_l,\quad
\boldsymbol{\alpha}=\sum_{l,k=1}^K a_{l,k}\psi_l\psi_k, \quad
\boldsymbol{\beta}=\sum_{l,k=1}^K b_{l,k}\psi_l\psi_k, \quad
\boldsymbol{\gamma}=\sum_{l,k=1}^K g_{l,k}\psi_l\psi_k,
\end{equation}
where $d_l$ forms a non-negative vector $\boldsymbol{D}=[d_1,\dots, d_K]^\top$ in $\mathbb{R}_+^K$, and $a_{l,k}$, $b_{l,k}$, and $g_{l,k}$ form non-negative matrices $\boldsymbol{A}=(a_{l,m})$, $\boldsymbol{B}=(b_{l,m})$, and $\boldsymbol{G}=(g_{l,m})$ in $\mathbb{R}_+^{K\times K}$. Therefore, the estimation of infinite-dimensional coefficient operators is transformed into the estimation of a finite-dimensional parameter vector:
\begin{equation*}
\theta=\text{vec}(\boldsymbol{D},\boldsymbol{A}, \boldsymbol{B}, \boldsymbol{G}) \in \Theta \equiv \mathbb{R}_+^{K+3\times K^2}.
\end{equation*}
Consequently, we will need to estimate the parameters $\theta = \text{vec}(\boldsymbol{D},\boldsymbol{A}_1,\boldsymbol{B}_1)$ for the models~\eqref{model-fgarch} when $p=q=1$.
The QMLE estimator in \cite{CFH+19} solves the optimisation
\begin{equation}\label{model-estimation}
\widehat{\theta}_N=\underset{\theta \in \boldsymbol{\Theta}}{\argmin}\frac{1}{N}\sum_{t=1}^N\sum_{l=1}^K\left\{ \frac{\left \langle \widetilde{y}_t^2,\psi_l \right \rangle}{\left \langle \widetilde{\sigma}_t^2,\psi_l \right \rangle}+\ln\left \langle \widetilde{\sigma}_t^2,\psi_l \right \rangle \right\}.
\end{equation}
Given the initial values $\widehat{\omega}(u)$ for $y_0(u)$ and a unit-valued constant function for $\sigma^2_0(u)$, respectively, the conditional variance $\sigma_t^2(u)$ in the model~\eqref{model-fgarchx} can be recursively calculated via
\begin{equation}\label{eq-recursive}
\widetilde{\sigma}_t^2(u)=\widetilde{\omega}(u)+\widetilde{\boldsymbol{\alpha}}(\widetilde{y}_{t-1}^{2})(u)+\widetilde{\boldsymbol{\beta}}(\sigma_{t-1}^2)(u)+\widetilde{\boldsymbol{\gamma}}(x_{t-1})(u). 
\end{equation}
The conditional variance $\widetilde{\sigma}_t^2(u)$ for model~\eqref{model-fgarch} can be deduced similarly. We rely on the regularity assumptions of \citet[][Theorem~2]{CFH+19} to guarantee that QMLE is a strongly consistent estimator of $\widehat{\theta}_N$.

Importantly, to make estimation feasible, we must decide on an optimal number for $K$ and the basis functions $\psi_l$. Existing methods include the use of goodness-of-fit tests \citep{Shibata81}, cross-validation \citep{BYZ10}, total variation explanation \citep{Cattell66}, and the eigenvalue ratio method \citep{LY12}. The first two methods are based on a grid search principle and can be computationally burdensome. The total variation explanation method determines $K$ by choosing an adequate number of basis functions to cumulatively explain a certain proportion of the variation, for example, 90\% of the total variations of our curve data. Lastly, the eigenvalue ratio method \citep{LY12, LRS20} selects $K = l$ via
\begin{equation}\label{eq-ratio}
    l = \arg \underset{1\leq l \leq \overline{L}}{\min} \frac{\widehat{\lambda}_{l+1}/\widehat{\lambda}_{1}}{\widehat{\lambda}_{l}/\widehat{\lambda}_{1}},
\end{equation}
where $\overline{L}$ is a prespecified positive integer and $\widehat{\lambda}_l$ is the $l$\textsuperscript{th} largest eigenvalue associated with the eigenfunction $\widehat{\phi}_l$. In this study, we use the eigenvalue ratio method. In choosing a suitable $\psi_l$, various types of data-driven methods are discussed in the literature. \citet{CFH+19} suggest using the truncated FPCA, while \citet{RWZ21} propose functional sparse and non-negative basis and truncated predictive factor methods to decompose $\{\widetilde{y}_t^2(u)\}$.

Note that in addition to FGARCH-type models, functional time series forecasting can also be performed using the FPCA approach of \cite{ANH15}, where the dynamics of functional scores are fitted and predicted by univariate or multivariate time series models. Simple adaptations for forecasting the conditional volatility of intraday return curves do not show clear empirical superiority in our unreported results. Therefore, we focus on FGARCH-type models, which benefit from the ability to capture the dynamics of the entire intraday return curves.

\subsection{Basis function specification}\label{sec-basis}

Based on the nature of the FX intraday return curves described in Section~\ref{sec-data}, it is clear that the models presented in Section~\ref{sec-model} struggle to capture stylised features to improve predictions of the conditional volatility of the OCIDR curves. In this section, we show that this useful information can still be used by selecting more nuanced basis functions $\psi_l$ in \eqref{model-coefficient}. As a crucial component of its dimension reduction procedure, basis selection drives the estimation and forecasting performance of FGARCH-type models. Usually, one can apply static \citep{CFH+19, PS23} or dynamic FPCA methods \citep{BYZ10, HKH15, SK22} to obtain basis functions. However, these methods overlook the rich cross-dependence of $\{y_t^2(u)\}$.

This section introduces two new data-driven approaches to decompose basis functions $\psi_l$ from the square OCIDR $\{y_t^2(u)\}$ curves. We decompose $\{y_t^2(u)\}$ here in the hope that the functional scores $\left< y^2_t(u), \psi_l(u)\right>$ preserve intraday conditional volatility dependence after we project the conditional volatility of $y_t(u)$ onto these basis functions. Note that intraday seasonal patterns are also absorbed into the estimated functional principal components and their associated scores. Therefore, modelling the dynamics of the scores considers intraday seasonality. 

\subsubsection{Multi-level FPCA}\label{sec-multi}

We first introduce a multi-level FPCA method to derive basis functions that account for cross-dependence between multiple FX intraday return curves. Instead of attempting to model cross-dependence dynamics directly through a multivariate FGARCH framework, which is still theoretically challenging, our approach captures and fits this stylised feature by incorporating the cross-dependence information through embedded data-driven basis functions in the dimensional reduction process.

As an extension of the functional mixed model \citep{MC06}, the multi-level functional time series model has been successfully applied across several fields, improving both exploration and forecasting through incorporating cross-dependence variations \citep[see, e.g.,][]{CF10, Shang16, SK22, SHX22}. Here, let us consider $d$-dimensional multivariate squared intraday return curves $\{y^2_{1,t}(u), y^2_{2,t}(u), \dots, y^2_{d,t}(u)\}$ ($d=3$ in our context), the multi-level functional data framework follows the idea of the common factor model by suggesting that these curves can be decomposed into an average functional mean $\mu_{(c)}(u)$, a forex-specific deviation from the averaged functional mean $\mu_{(j)}(u)$, a common trend across forex rates $U_{(c),t}(u)$, and a forex-specific residual trend $U_{(j),t}(u)$, for $1\leq j \leq d$. Thus, the $j$\textsuperscript{th} squared OCIDR curve can be written as,
\begin{equation*}
y^2_{j,t}(u) =\mu_{(c)}(u)+\mu_{(j)}(u)+U_{(c),t}(u)+U_{(j),t}(u),
\end{equation*}
where each term is practically estimated via:
\begin{equation}
\begin{split}
& \widehat{\mu}_{(c)}(u) = \frac{1}{Nd}\sum_{t=1}^{N}\sum_{j=1}^{d} y^2_{j,t}(u), \\
& \widehat{\mu}_{(j)}(u) = \frac{1}{N}\sum_{t=1}^{N} y^2_{j,t}(u) - \widehat{\mu}_{(c)}(u),\\
& \widehat{U}_{(c),t}(u)=\frac{1}{d}\sum_{j=1}^{d}y^2_{j,t}(u)-\widehat{\mu}_{(c)}(u), \\
& \widehat{U}_{(j),t}(u) = y^2_{j,t}(u)-\widehat{\mu}_{(c)}(u)-\widehat{\mu}_{(j)}(u)-\widehat{U}_{(c),t}(u).
\end{split}
\end{equation}

Concentrating on the common and forex-specific variation components $\{U_{(c),t}(u)\}$ and $\{U_{(j),t}(u)\}$, we can estimate their long-run covariance operators $c_{U_c,N}(u,v)$ and $c_{U_j,N}(u,v)$, respectively. Linearly independent basis functions $\{\widehat{\psi}^{(c)}_1(u), \dots, \widehat{\psi}^{(c)}_{K^c}(u) \}$ and $\{ \widehat{\psi}^{(j)}_1(u), \dots, \widehat{\psi}^{(j)}_{K^j}(u)\}$ associated with the corresponding eigenvalues can be obtained by eigen-decomposing the uncorrelated operators $c_{U_c,N}(u,v)$ and $c_{U_j,N}(u,v)$, respectively. As we explained in Section~\ref{sec-model}, in practice, the values of dimensions $K^c$ and $K^j$ can be chosen using the eigenvalue ratio method of \cite{LRS20}.

Due to having two types of basis functions representing common variations and forex-specific variations, we project the FGARCH-type models onto $\{\widehat{\psi}^{(c)}_1(u), \dots, \widehat{\psi}^{(c)}_{K^c}(u) \}$ and $\{ \widehat{\psi}^{(j)}_1(u), \dots, \widehat{\psi}^{(j)}_{K^j}(u)\}$ and then estimate the coefficients in models~\eqref{model-fgarch} and~\eqref{model-fgarchx}, respectively. For instance, in terms of the FGARCH-X model, the coefficients of the conditional volatility equation in~\eqref{model-coefficient} can then be estimated through,
\begin{equation}\label{model-multicoefficient}
\begin{split}
& \omega^{(j)}=\sum_{l_1=1}^{K^c} d_{l_1} \psi^{(c)}_{l_1} + \sum_{l_2=1}^{K^j} d_{l_2} \psi^{(j)}_{l_2}  ,\\
& \boldsymbol{\alpha}^{(j)}=\sum_{l_1,k_1=1}^{K^c} a_{l_1,k_1}\psi^{(c)}_{l_1}\psi^{(c)}_{k_1} + \sum_{l_2,k_2=1}^{K^j} a_{l_2,k_2}\psi^{(j)}_{l_2}\psi^{(j)}_{k_2}\\
& \boldsymbol{\beta}^{(j)}=\sum_{l_1,k_1=1}^{K^c} b_{l_1,k_1}\psi^{(c)}_{l_1}\psi^{(c)}_{k_1} + \sum_{l_2,k_2=1}^{K^j} b_{l_2,k_2}\psi^{(j)}_{l_2}\psi^{(j)}_{k_2},\\
& \boldsymbol{\gamma}^{(j)}=\sum_{l_1,k_1=1}^{K^c} g_{l_1,k_1}\psi^{(c)}_{l_1}\psi^{(c)}_{k_1} + \sum_{l_2,k_2=1}^{K^j} g_{l_2,k_2}\psi^{(j)}_{l_2}\psi^{(j)}_{k_2}.
\end{split}
\end{equation}
The estimated kernel coefficients comprise a common variation component captured by the bases $\psi^{(c)}$ and a forex-specific component captured by the bases $\psi^{(j)}$.

\subsubsection{Long-range dependent FPCA}\label{sec-long}

Treating $\{y_t^2{(u)}\}$ as short-range dependent curves can sometimes be inappropriate, as we learnt in Table~\ref{table-data} that intraday return curves are long-range conditional heteroscedastic. In this section, we introduce a long-range dependent FPCA (LFPCA) method to obtain embedded basis functions with a long-range dependence structure.

Inspired by \citet{LRS20}, we write $\{y_t^2(u)\}$ as a functional linear process,
\begin{equation}\label{eq-flp}
y_t^2(u)=\sum_{j=0}^\infty b_j \eta_{t-j}(u),
\end{equation}
where $\{b_j, j\geq 0\}$ is a sequence of coefficients and $\eta_{t-j}(u) \in \mathcal{L}^2[0,1]$ represents a series of independent identically distributed random functions. Under stationarity, we say that $\{y_t^2(u)\}$ is short-range dependent if $\sum_{j=0}^\infty |b_j| <\infty$; and $\{y_t^2(u)\}$ is long-range dependent if the summation is unbounded. We denote the memory parameter $a$, which describes the dependence strength of $\{y_t^{(2)}\}$ by reflecting $b_j \sim j^{a-1}$ in~\eqref{eq-flp}. Specifically, $\{y_t^{(2)}\}$ is deemed to be a stationary long-range dependent curve sequence when $1/2\leq a<1$; and it becomes non-stationary when $a\geq 1$. 

\begin{figure}[H]
\centering
\caption{$P$-values of stationarity tests for the squared Overnight Cumulative Intraday Return (OCIDR) curve, $\{y_t^2(u)\}$, across the various training samples utilised in the forecasting exercise in Section~\ref{sec-app}. $TSample\_x$ indicates the results for each training sample with $x$ ranging from 1 to 19. For example, the first training sample $TSample\_1$ covers the period 07-January-2014 to 17-May-2017 with the periods covered by the other training samples given in footnote~\ref{train_dates}. P-values less than the outlined 5\% threshold indicate non-stationarity for that training sample.}
\label{figure-ocidr-stationary}
\includegraphics[width=14cm]{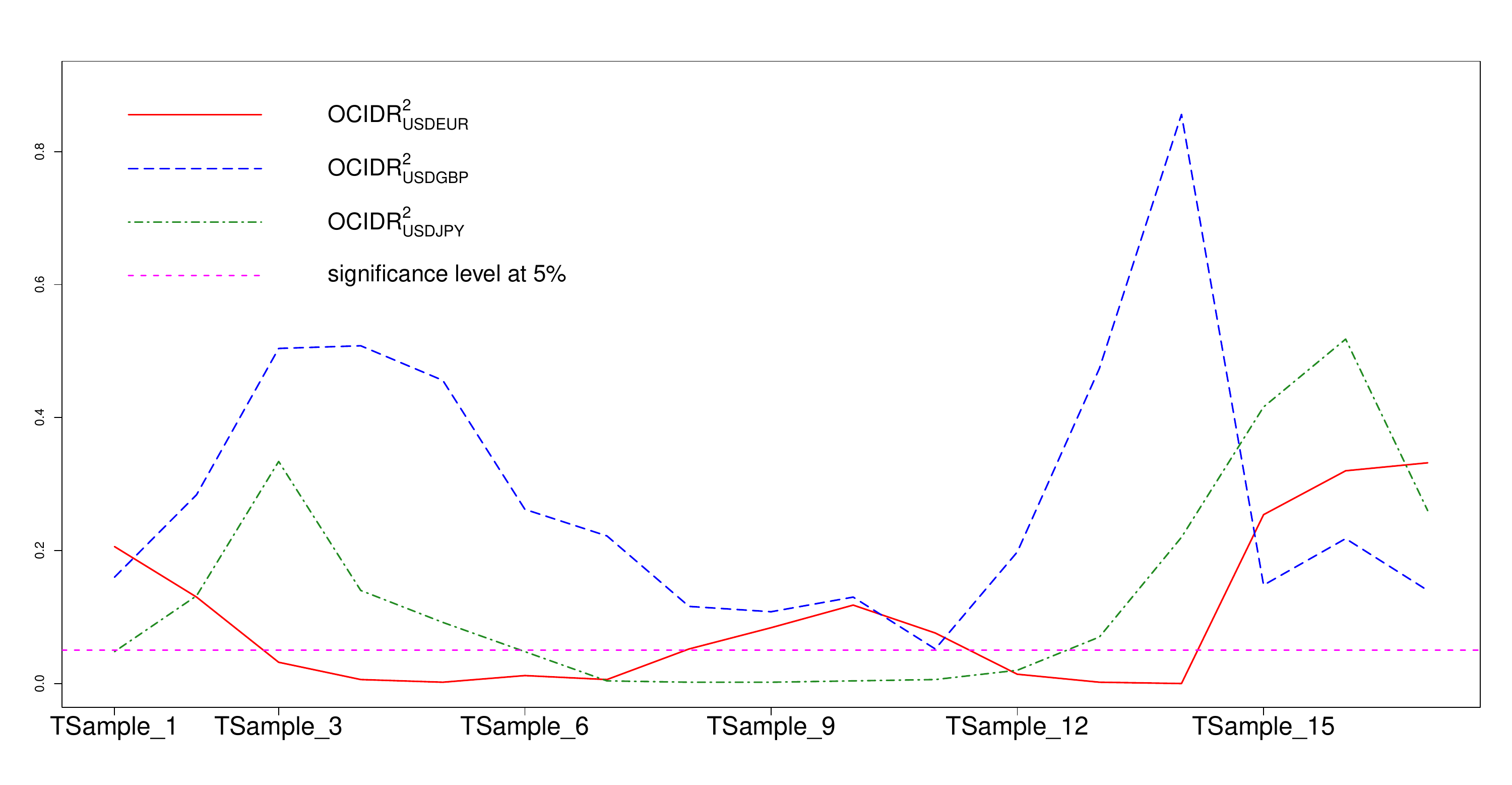}
\end{figure}

In addition, we empirically test the stationarity of $\{y_t^2(u)\}$ across the multiple forecasting training samples that we employ in Section~\ref{sec-app}. Figure~\ref{figure-ocidr-stationary} shows the \textit{p}-values of stationary tests implemented for the squared OCIDR curves in 19 training samples.\footnote{The periods covered by 19 training samples, which are updated every two months: 07-January-2014 to 17-May-2017; 27-March-2014 to 21-July-2017; 13-June-2014 to 22-September-2017; 28-August-2014 to 24-November-2017; 18-November-2014 to 31-January-2018; 10-February-2015 to 04-April-2018; 07-May-2015 to 06-June-2018; 24-July-2015 to 08-August-2018; 20-October-2015 to 10-October-2018; 05-January-2016 to 12-December-2018; 15-March-2016 to 26-February-2019; 19-May-2016 to 30-April-2019; 22-July-2016 to 02-July-2019; 23-September-2016 to 03-September-2019; 29-November-2016 to 05-November-2019; 01-February-2017 to 22-January-2020; 05-April-2017 to 25-March-2020; 13-June-2017 to 27-May-2020, 15-August-2017 to 11-August-2020. \label{train_dates}} The results indicate that $\{y_t^2(u)\}$ could be stationary or non-stationary, such that the $\{y_t^2(u)\}$ samples of USD/GBP are stationary, while the other two currency pairs exhibit non-stationarity. In the following, we detail the LFPCA approaches to derive data-driven bases under these two scenarios.

\paragraph{Stationary long-range dependent curves}

Following \citet{LRS20}, the curve time series are stationary long-range dependent if the kernel coefficient operators in~\eqref{eq-flp} are not square-summable and $1/2 \leq a<1$. Let the unnormalised long-run covariance function be:
\begin{equation}\label{eq-ulrc}
c_N(u,v) = \mathbb{E}\left[\sum_{t=1}^N\sum_{s=1}^N y_t^2(u) y_s^2(v)\right].
\end{equation}
According to Assumption~2 and the large-sample properties in \citet{LRS20}, we learn that the long-range dependent $y_{t}^{2}(u)$ can be decomposed into two parts: 
\begin{inparaenum}
\item[1)] the projection of $y_t^2(u)$ onto a finite number of sub-spaces that capture the long-range dependence structure of $y_t^2(u)$; and 
\item[2)] the projection onto the remaining infinite-dimensional sub-spaces stores the short-range dependence. 
\end{inparaenum}
Thus, the long-range dependence of $y_t^2(u)$ can be captured by projecting onto a dominant sub-space $\mathcal{S}_{1}$ spanned over $\{\psi_1, \psi_2, \dots, \psi_{K_1}\}$. Then, dimension reduction can be achieved via the FPCA method on the normalised long-run covariance:
\begin{equation*}
\widetilde{c}_N(u,v) = \underset{N\rightarrow \infty}{\lim} \frac{1}{N^{3-2a_1}}c_N(u,v),
\end{equation*}
where $a_1$ is a memory parameter that measures the dependence strength of the signal for the projection of $y_t^2(u)$ on $\mathcal{S}_{1}$. The normalised long-run covariance $\widetilde{c}(u,v)_N$ is not observable, so it has to be estimated through,
\begin{equation}\label{eq-normallr1}
\widetilde{c}_{h,N}(u,v) = \frac{1}{h^{3-2a_1}}\sum_{|\ell|\leq h} (h-|\ell|) \widehat{r}_\ell(u,v),
\end{equation}
where $\widehat{r}_\ell(u,v)$ has been defined as,
\begin{equation}\label{eq-longruncov}
    \widehat{r}_\ell(u,v) = \left\{\begin{matrix}
\frac{1}{N} \sum_{t=1}^{N-\ell}[y_t^2(u)-\overline{y}_N^2(u)][y_{t+\ell}^2(v)-\overline{y}_N^2(v)], \mbox{ } \ell\geq0.
 \\
\frac{1}{N} \sum_{t=1}^{N-|\ell|}[y_{t+|\ell|}^2(u)-\overline{y}_N^2(u)][y_t^2(v)-\overline{y}_N^2(v)], \mbox{ } \ell<0.
\end{matrix}\right.
\end{equation}
The bandwidth $h$ is chosen using the plug-in algorithm proposed in \citet{RS17} to minimise the estimator's asymptotic mean-squared normed error. One can use the eigenvalue ratio method to determine the dimension $K_1$ of the dominant sub-space. \citet{LRS21} propose a semi-parametric feasible local Whittle estimator $\widehat{a}_1$ for the memory parameter in the context of functional time series. We suppress the technical details to save some space, and interested readers can refer to \citet{LRS21}. Consequently, we obtain the consistent estimator of $\widetilde{c}_h(u,v)_N$ by substituting $\widehat{a}_1$ into~\eqref{eq-normallr1}, with long-range dependent basis functions $\widehat{\psi}_l$ accessed through the eigen-decomposition method.

\paragraph{Non-stationary long-range dependent curves}

We now discuss the case where $\{y_t^2(u)\}$ constitutes a non-stationary curve sequence (when $a> 1/2$). \citet{CKP16} is the seminal work in this field, paving a foundation for the subsequent developments in cointegration \citep{NSS23, SS23}. They decompose a non-stationary functional time series from infinite-dimensional Hilbert space into a finite-dimensional I(1) subspace and infinite stationarity subspace. The main focus of the literature then turns to determine the finite dimension of the non-stationarity subspace \citep[see, e.g.,][]{LRS20ns}. In this paper, we apply the approach proposed in \citet{LRS20ns}, which assumes the data of interest follow a non-stationary fractionally integrated functional process as specified in model~\eqref{eq-flp}. This is a preferred method as it allows an unknown order of integration, which can be estimated using a feasible local Whittle estimator as in the stationary case. Furthermore, similar to the stationary long-range dependent case, the linear operator $b_j$ is not square-summable. However, the memory parameter can be greater than one under non-stationarity. \citet{LRS20ns} show that such non-stationary functional curves can be divided into a finite dominant subspace $\overline{\mathcal{S}}_1$ that retains the non-stationary sample information and the remaining infinite asymptotically negligible subspaces. According to Theorem 3.2 in \citet{LRS20ns}, we can define the normalised sample covariance operator on $\overline{\mathcal{S}}_1$,
\begin{equation}\label{eq-normallr2}
\overline{c}_N(u,v)= \frac{1}{N^{2a_1} L_1^2(N)}\sum_{t=1}^{N} y_t^2(u)\otimes y_t^2(v),
\end{equation}
where $\otimes$ denotes the tensor product. For the estimation of the memory parameter $a_1$, we refer to the local Whittle estimator in \citet{LRS20ns}. By eigen-decomposing $\overline{c}_N(u,v)$, we obtain $\widehat{\psi}_{l}$ eigenfunctions, for $1\leq l \leq \overline{K}_1$ , the span of which can generate the dominant space $\overline{\mathcal{S}}_1$. Similar to the stationary case, $\overline{K}_1$ can be determined using the eigenvalue ratio method.

Notably, we need to estimate the memory parameter $a_1$ for both normalised covariance operators in~\eqref{eq-normallr1} and~\eqref{eq-normallr2}. In fact, it is not difficult to see that $\widetilde{c}_{h,N}(u,v)$ and $\overline{c}_N(u,v)$ are proportional to $\widetilde{c}^*_N(u,v)$, meaning that the empirical principal components decomposed from both fronts are the same as the empirical principal components decomposed from the latter. This allows us to eliminate the effect of $a_1$. Therefore, we eigen-decompose $\widetilde{c}^*_N(u,v)$ and obtain functional principal components $\{\widehat{\psi}_1, \dots, \widehat{\psi}_{K_1}\}$ for the basis functions in either stationary or non-stationary long-range dependent cases.

\section{Predictive accuracy}\label{sec-app}

We now perform a conditional volatility forecasting evaluation for our FX rate intraday returns. The data sample is initially partitioned into a training sample (07-January-2014 to 17-May-2017) and a forecasting sample (18-May-2017 to 30-September-2020). A rolling window approach is applied to predict the one-step-ahead conditional volatility of the FX OCIDR curves. To adapt to changing market states, the forecasting model is estimated every week to enable more accurate predictions. The initial training sample covers 07-January-2014 to 17-May-2017, and the last training sample is from 27-September-2017 to 24-September-2020. To this end, we form 169 different training samples to forecast the holdout data in the forecasting period.

We compare the forecasting models by assessing the accuracy of intraday and inter-daily volatility forecasts. The latter is feasible because the closing points of the intraday volatility forecasts for the OCIDR curves form a sequence of conditional volatility forecasts for daily returns, as we emphasised in Section~\ref{sec-data}. In order to obtain the forecasting errors of the model candidates, ideally, we need unbiased and consistent estimators of latent volatility. This is feasible when we assess the inter-daily volatility forecasts, as the realised volatility $\text{RV}_t$ is an unbiased and consistent estimator that has been widely used in the literature \citep{BS02, HL2006}. Incorporating the overnight effect, we can calculate the realised volatility by summing squared intraday log returns with the squared overnight return,
\begin{equation*}
\text{RV}_{t} = \sum_{j=1}^J\{\ln P_t(j\cdot \Delta) - \ln P_t [(j-1)\cdot \Delta]\}^2+[\ln P_t(0)-\ln P_{t-1}(1)]^2,
\end{equation*}
where $\Delta$ denotes an intraday sampling frequency. However, there is still no consistent and unbiased estimator for intraday volatility. We then consider the actual values of intraday volatility using the proxies of the squared OCIDR curves. We are aware that substituting the unknown consistent estimator of intraday conditional variance with squared OCIDR curves may result in unreliable and inferior results, similar to the same issue for scalar GARCH-type models discussed by \citet{HL2006} and for multivariate cases discussed by \citet{PS2009}. 

Nonetheless, it is tractable to evaluate forecasting performance with noisy proxies. \citet{P2011} documented that a noisy proxy can still be used to evaluate the volatility forecasts if a ``robust" loss function is considered, which maintains an invariant property of ranking forecasts using an imperfect volatility proxy. Following \citet{P2011}, we apply two robust loss functions, namely mean squared forecasting error (MSFE) and QLIKE. Unlike MSFE, QLIKE is asymmetric; it penalises large under-forecasts much heavier than over-forecasts. Below, we define MSFE and QLIKE loss functions. For inter-daily conditional volatility forecasts, we have
\begin{equation}\label{eq-forecasterror-daily}
\begin{split}
\mbox{MSFE: }& L(\widetilde{\sigma}_t^2, \widehat{\sigma}_t^2) = \frac{1}{T}\sum_{t=1}^{T}[(\widetilde{\sigma}_t^2-\widehat{\sigma}_t^2)^2],\\
\mbox{QLIKE: }& L(\widetilde{\sigma}_t^2, \widehat{\sigma}_t^2) = \frac{1}{T}\sum_{t=1}^{T}\left[\log(\widehat{\sigma}^2_t) + \frac{\widetilde{\sigma}^2_t}{\widehat{\sigma}^2_t} \right],
\end{split}
\end{equation}
where $\widetilde{\sigma}_t^2$ is the realised volatility and $\widehat{\sigma}_t^2$ is the volatility forecast. The loss functions for intraday volatility curves can be similarly defined as
\begin{equation}\label{eq-forecasterror-intraday}
\begin{split}
\mbox{MSFE: }& L[\widetilde{\sigma}_t^2(u), \widehat{\sigma}_t^2(u)] = \frac{1}{T}\sum_{t=1}^{T}\left\{\frac{1}{J}\sum_{j=1}^{J}\left[\widetilde{\sigma}_t^2(u_j)-\widehat{\sigma}_t^2(u_j)\right]^2\right\},\\
\mbox{QLIKE: }& L[\widetilde{\sigma}_t^2(u), \widehat{\sigma}_t^2(u)] = \frac{1}{T}\sum_{t=1}^{T}\left\{\frac{1}{J}\sum_{j=1}^{J} \left[ \log \widehat{\sigma}^2_t(u_j) + \frac{\widetilde{\sigma}^2_t(u_j)}{\widehat{\sigma}^2_t(u_j)} \right]\right\},\\
\end{split}
\end{equation}
where operations are performed in a pointwise manner, that is, the proxy of the squared OCIDR curves $\widetilde{\sigma}_t^2(u)$ and the predicted volatility curves $\widehat{\sigma}^{2}_{t}(u)$ are measured on the $J$ grid points of $\widetilde{\sigma}_{t}^{2}(u_j)$ and $\widehat{\sigma}^{2}_{t}(u_j)$. 

\subsection{Comparison of FGARCH-type models}

First, we assess the predicability of FGARCH-type models. In total, we consider eight forecasting models that are the permutations of two model specifications. These are FGARCH$(1,1)$ and the FGARCH-X model, along with four dimension reduction basis functions: functional principal components (FPCA), dynamic functional principal components (DFPCA) of \cite{SK22}, long-range dependent functional principal component (LFPCA), and multi-level functional principal components (MFPCA). We demean the OCIDR curves to obtain $\widetilde{y}_t(u)$ by removing the functional mean. The parameters $K$ in~\eqref{model-coefficient} and $K^c$ and $K^j$ in~\eqref{model-multicoefficient}, that is, the dimension of our basis functions, have to be determined when estimating the forecasting models. We select the appropriate number of these parameters using the eigenvalue ratio method~\eqref{eq-ratio} for each model estimation. This method indicates that using only the first functional principal component, that is, Basis function 1, is adequate in all cases. To inform our parameter selection, we decompose the data-driven basis functions from squared intraday returns. Table~\ref{table-basis} shows that the first basis explains most of the total variation. The first basis of DFPCA roughly accounts for 80\%, with the first basis generally explaining 95\% for the other approaches. The results are produced using the first training sample (07-January-2014 to 17-May-2017), and the observations generalise when adopting other training samples. In addition, by testing the goodness-of-fit of the FGARCH$(1,1)$ and FGARCH-X models based on the chosen data-driven bases, we learn that the models are adequate to explain the conditional heteroskedasticity of the FX OCIDR curves. The results of the goodness-of-fit test by \cite{RWZ20} are omitted here to save space, but can be obtained upon request.
\begin{table}[H]
\centering
\caption{The percentage of total variation explained by the decomposed basis functions for truncated functional principal components (TFPCA), dynamic functional principal components (DFPCA), long-range dependent functional principal components (LFPCA), and multi-level functional principal components (MFPCA). COM and RES represent the common and forex-specific variation components of the MFPCA basis functions as presented in~\eqref{model-multicoefficient}. Our first training sample of 07-January-2014 to 17-May-2017 is specified.}\label{table-basis}
\begin{adjustbox}{max width=\linewidth}
\begin{tabular}{@{}cccccccccccccccc@{}}
\toprule
& \multicolumn{5}{c}{USD/EUR}                         & \multicolumn{5}{c}{USD/GBP}                         & \multicolumn{5}{c}{USD/JPY}       
\\\cmidrule{2-16}
& TFPCA  & DFPCA & LFPCA & \multicolumn{2}{c}{MFPCA} & TFPCA  & DFPCA & LFPCA & \multicolumn{2}{c}{MFPCA} & TFPCA  & DFPCA & LFPCA & \multicolumn{2}{c}{MFPCA} \\\midrule
&       &       &       & COM          & RES         &       &       &       & COM          & RES         &       &       &       & COM          & RES         \\\midrule
Basis~1 &  0.853  &  0.788  &  0.988  &     0.772    &     0.849    & 0.998      &    0.924    &   0.968    &    0.772       &   0.817    &    0.819   &  0.654     &    0.963    &    0.772      &    0.784  \\
Basis~2 &  0.069  &  0.092  &  0.008  &    0.186     &    0.108    &  0.001     &  0.055     &   0.026    &     0.186      &   0.148    &   0.124    &  0.206     &   0.023     &    0.186      &  0.151  \\
Basis~3 &  0.036  &  0.058  &  0.002  &     0.021    &    0.025    &   0.001    &  0.010     &   0.004    &      0.021     &   0.017    &   0.039    &  0.054     &  0.008      &    0.021   &  0.032   \\
Basis~4  &  0.021  &  0.025  &  0.001  &    0.006     &    0.006    &  0.000     &   0.005    &    0.001   &     0.006      &    0.007   &    0.010   &  0.033     &   0.002     &     0.006     &   0.011  \\
Basis~5 &  0.009  &  0.011  &  0.000 &     0.004    &   0.003     &    0.000   &  0.002     &   0.000    &    0.004       &    0.003   &    0.003   &  0.022     &   0.001     &   0.004       &  0.005 \\\bottomrule
\end{tabular}
\end{adjustbox}
\end{table}

\begin{figure}[H]
\centering
\caption{Plots of pointwise intraday volatility forecasts and squared OCIDR curves of USD-EUR over 24-hour trading sessions obtained from FGARCH and FGARCH-X models with the basis functions are derived by using the MFPCA approach. Our first training sample from 7 January 2014 to 17 May 2017 and a forecasting sample from 18 May 2017 to 30 September 2020 are specified with a rolling forecast window dynamically updated every week. The left sub-figure displays the plots of mean curves, and the right sub-figure shows the plots of median curves.}\label{figure-intravol}
\includegraphics[width=18cm]{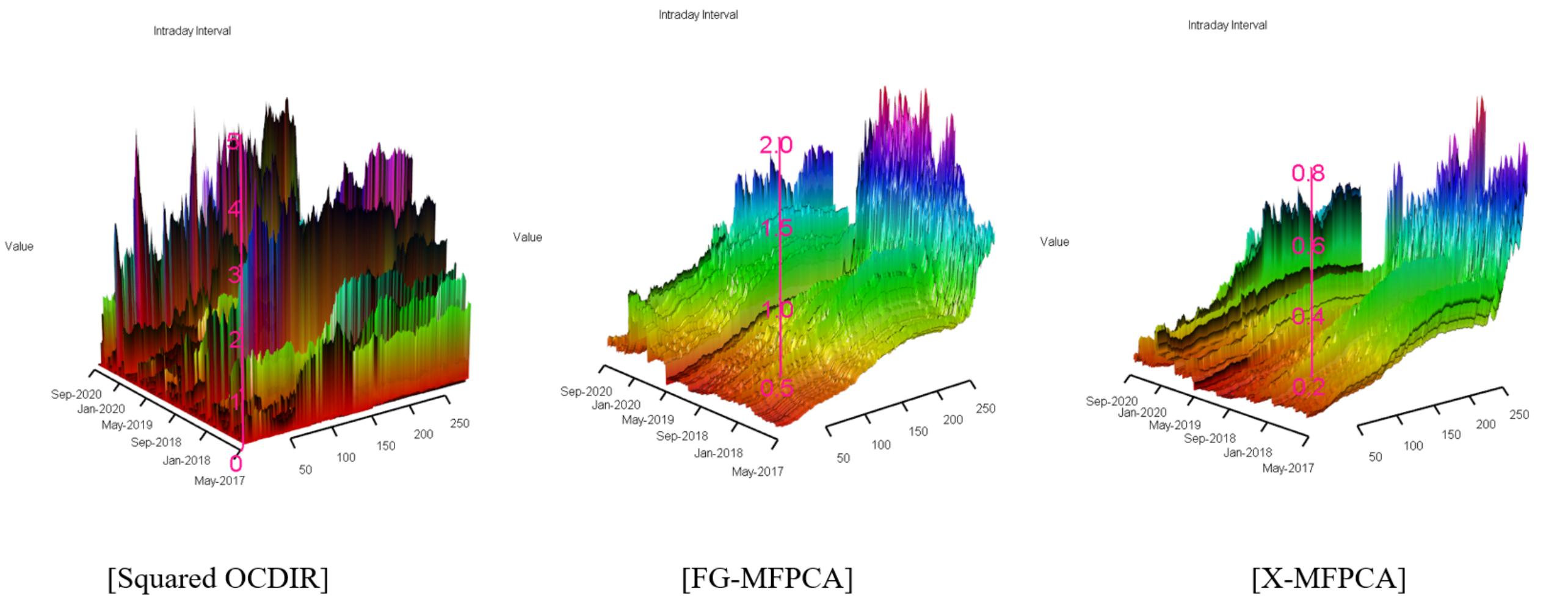}
\end{figure}

Figure~\ref{figure-intravol}  displays the intraday volatility forecasts of USD/EUR obtained from the FGARCH and FGARCH-X models using MFPCA bases, as well as the squared OCIDR curves. We find that both the FGARCH and FGARCH-X models reasonably attempt to represent the dynamics of squared OCIDR curves, forecasting intraday volatility to increase toward the end of the trading day. This is because the bases derived from the training sample capture the same pattern, with cumulative intraday returns generally increasing as the day progresses, which is also reflected in the squared OCIDR curves. MFPCA-embedded model captures more spikes at the end of the trading day, the FGARCH-X models in average provide forecasts closer to the squared OCIDR curves. Additionally, intraday volatility forecasts are notably higher in 2018 and 2020, capturing features similar to those of squared OCIDR curves. This increased volatility corresponds to multiple US interest rate hikes, the US-China trade war, political uncertainty in Europe in 2018, and the outbreak of the COVID-19 pandemic that led to significant central bank interventions in 2020. In addition, after accounting for the bid-ask spread, the FGARCH-X model forecasts volatility on a relatively lower scale, although the overall pattern remains similar to that predicted by the FGARCH model.


Table~\ref{table-forecast-error} shows the loss functions based on our forecasting errors. The comparative performances are mostly consistent when measured using either MSFE or QLIKE loss functions. Notably, in both Panels~A and~B, the models embedded with MFPCA bases (which account for cross-dependence among FX rates) generally produce the smallest forecasting errors for USD/EUR and USD/JPY. This evidences that our multivariate approach disentangles common variation from currency-specific information, capturing joint shocks for volatility modelling more effectively. The results for USD/GBP are somewhat mixed, with the smaller MSFE suggesting the use of TFPCA or LFPCA bases. We also notice that MFPCA bases are more effective in improving the accuracy of the FGARCH(1,1) model compared to the FGARCH-X model. This may be because the cross-dependence variations driven by the MFPCA bases do not properly consider the variations in the covariate~X. Moreover, the models using an LFPCA basis, particularly in Panel B, outperform the TFPCA and DFPCA bases models, demonstrating the benefit of explaining the long-range dependence of the OCIDR curves. The models relying on a DFPCA basis perform slightly better than the TFPCA counterparts in forecasting intraday and inter-daily conditional volatility. Furthermore, the MSFE and QLIKE generally give consistent results when assessing the inter-daily conditional volatility forecasts. The results are more mixed in the USD/GBP intraday conditional volatility forecasts, where some inconsistency arises from specifying the TFPCA or LFPCA bases. Lastly, to compare the FGARCH and FGARCH-X models, we observe that the FGARCH-X models generate considerably lower forecasting errors than the FGARCH model in both the intraday and inter-daily conditional volatility forecasts. This is because different FPCA methods serve as data-driven approaches that capture only the dynamics of the squared OCIDR process. By incorporating bid-ask spread information, the FGARCH-X model offers the opportunity to correct the level of conditional volatility modelling, enhancing its predictive accuracy.
\begin{table}[H]
\tabcolsep 0.2in
\renewcommand{\arraystretch}{1.05}
\begin{center}
\caption{One-step-ahead forecasting errors of FGARCH and FGARCH-X models with four types of data-driven basis functions: truncated functional principal components (TFPCA), dynamic functional principal components (DFPCA), long-range dependent functional principal components (LFPCA), and multi-level functional principal components (MFPCA). Panel A documents the MSFE and QLIKE forecasting errors of intraday conditional volatility as calculated using Equation~\eqref{eq-forecasterror-intraday}, with Panel B showing the forecasting errors of inter-daily conditional volatility as calculated using Equation~\eqref{eq-forecasterror-daily}. Our first training sample of 07-January-2014 to 17-May-2017 and a forecasting sample of 18-May-2017 to 30-September-2020 are specified with a rolling forecast window that is dynamically updated weekly.}\label{table-forecast-error}
\begin{adjustbox}{max width=\linewidth}
\begin{tabular}{@{}lccccccc@{}}
\toprule
&       & \multicolumn{2}{c}{USD/EUR} & \multicolumn{2}{c}{USD/GBP} & \multicolumn{2}{c}{USD/JPY} \\\cmidrule{2-8}
&  Method     & MSFE         & QLIKE        & MSFE         & QLIKE        & MSFE         & QLIKE       \\\midrule
&       \multicolumn{7}{c}{Panel~A: Intraday conditional volatility forecasts}                        \\\midrule
\multirow{4}{*}{FGARCH} & TFPCA & 0.4002 & -0.2744 & \bf{0.8198} & 1.7702 & 0.5556 & -0.4632 \\
& DFPCA & 0.3990 & -0.2838 & 0.8353 & -0.0055 & 0.5517 & -0.4882 \\ 
& LFPCA & 0.3944 & -0.2698 & 0.8411 & \bf{-0.2386} & 0.5459 & 7.8980 \\
& MFPCA & \bf{0.3798} & \bf{-0.3652} &  0.8425  &  -0.2174  & \bf{0.5329} & \bf{-0.5516} \\
\\
\multirow{4}{*}{FGARCH-X}  & TFPCA & 0.1330 & -0.8434 & 0.7055 & 2.4777 & 0.4684 & -0.7467 \\
& DFPCA & 0.1330 & -0.8426 & 0.6980 & 0.1923 & 0.4664 & -0.7788 \\
& LFPCA & 0.1333 & -0.8289 & \bf{0.6904} & \bf{-0.2610} & 0.4620 & 1.3790 \\
& MFPCA & \bf{0.1323} & \bf{-0.8641} & 0.6948 & -0.1467 & \bf{0.4468} & \bf{-0.8077} \\  \midrule
&       \multicolumn{7}{c}{Panel~B: Inter-daily conditional volatility forecasts}                     \\\midrule
\multirow{4}{*}{FGARCH} & TFPCA & 0.5134 & 0.1162 & \bf{0.3375} & 2.3640 & 0.8573 & -0.1702 \\
& DFPCA & 0.4952 & 0.1039 & 0.3655 & 0.5769 & 0.8538 & -0.1729 \\
& LFPCA & 0.4564 & 0.0763 & 0.3471 & \bf{0.0853} & 0.8096 & 3.6107 \\
& MFPCA & \bf{0.3398} & \bf{-0.0435} & 0.3528 & 0.1034 & \bf{0.7900} & \bf{-0.2572} \\
\\
\multirow{4}{*}{FGARCH-X}  & TFPCA & 0.0804 & -0.4677 & 0.2456 & 3.1062 & 0.7250 & -0.4982 \\
& DFPCA & 0.0807 & -0.4700 & 0.2286 & 0.8918 & 0.7231 & -0.5031 \\
& LFPCA & 0.0781 & -0.4855 & \bf{0.2084} & \bf{0.0830} & 0.7228 & \bf{-0.5045} \\
& MFPCA & \bf{0.0740} & \bf{-0.5155} & 0.2293 & 0.1787 & \bf{0.7099} & -0.4847 \\
\bottomrule
\end{tabular}
\end{adjustbox}
\end{center}
\end{table}

We then apply the Model Confidence Set (MCS) of \cite{HLN11} and the \citeauthor{DM1995}'s \citeyearpar{DM1995} test to assess the relative performances in multi-model and pairwise models, respectively. Table~\ref{table-mcs} shows the forecasting models designated to lie in the superior set of models at 95\% significance level for each FX rate. Our findings indicate that the intraday bid-ask spread information significantly improves predictability by reducing forecasting errors. The FGARCH-X model is always elected to be the superior model to forecast either intraday conditional volatility in Panel~A or inter-daily conditional volatility in Panel~B. 
\begin{table}[H]
\tabcolsep 0.2in
\centering
\caption{Results of model confidence set tests \citep{HLN11} based on the MSFE and QLIKE forecasting errors presented in Equations~\eqref{eq-forecasterror-intraday} and~\eqref{eq-forecasterror-daily} at 95\% confidence set level. FGARCH labelled FG-, and FGARCH-X, labelled X-, models are specified with four types of data-driven basis functions: truncated functional principal components (TFPCA), dynamic functional principal components (DFPCA), long-range dependent functional principal components (LFPCA), and multi-level functional principal components (MFPCA). Our first training sample of 07-January-2014 to 17-May-2017 and a forecasting sample of 18-May-2017 to 30-September-2020 are specified with a rolling forecast window dynamically updated weekly. Panel A displays the models falling into the confidence sets that forecast intraday conditional volatility. Panel~B shows the models that fall into the confidence sets that forecast inter-daily conditional volatility.}\label{table-mcs}
\begin{adjustbox}{max width=\linewidth}
\begin{tabular}{@{}lrrrrrr@{}}
\toprule
			& \multicolumn{2}{c}{USD/EUR}                                             & \multicolumn{2}{c}{USD/GBP}                                                                                                & \multicolumn{2}{c}{USD/JPY}                                              \\\cmidrule{2-7}
			& MSFE                                                        & QLIKE    & MSFE                                                       & QLIKE                                                        & MSFE     & QLIKE                                                        \\\cmidrule{2-7}
			& \multicolumn{6}{c}{Panel A: Intraday conditional volatility forecasts}                       \\\midrule
			95\% & X-MFPCA                                                    & X-MFPCA & X-TFPCA                                                   & X-LFPCA                                                     & X-MFPCA &  X-MFPCA \\\midrule
			& \multicolumn{6}{c}{Panel B: Inter-daily conditional volatility forecasts}                                              \\\midrule
			95\% & X-MFPCA                                             &  X-MFPCA & X-TFPCA  & \begin{tabular}[c]{@{}c@{}}FG-LFPCA\\ X-LFPCA\end{tabular}                                                & X-MFPCA & X-LFPCA\\\bottomrule
\end{tabular}
\end{adjustbox}
\end{table}

Our findings complement the literature and provide further evidence that intraday bid-ask spread information provides predictability to conditional volatility in FX markets. In addition, to compare the different basis functions used, the models associated with the MFPCA approach are still chosen most often for USD/EUR and USD/JPY. For the USD/GBP rate, the FGARCH-X model using the LFPCA basis mostly falls into all confidence sets, with the exception that the TFPCA is selected when the QLIKE measurements are used, which follows the results of Table~\ref{table-forecast-error}. Furthermore, it is interesting to note that, according to Table~\ref{table-data}, USD/GBP is more long-range dependent compared to the other two rates at the second moment with the long-memory parameter $\widehat{a}_{x^2}=0.33$, which could be the reason why the models associated with the LFPCA basis outperform for USD/GBP.

Table~\ref{table-dm} shows the results of the Diebold-Mariano test on pairwise forecasting models. Twelve model pairs are compared, including ten pairs of the FGARCH and FGARCH-X models, with TFPCA and DFPCA serving as benchmarks. Additionally, two pairs of FGARCH and FGARCH-X models are compared, where the same types of LFPCA and MFPCA bases are applied. In terms of intraday conditional volatility forecasts, the results demonstrate that, in a few cases for all FX rates, the benchmark models significantly underperform the model, relying on the other proposed bases. The MFPCA, as the superior basis, consistently predicts more accurate conditional volatility for FGARCH models in USD/EUR. However, the relative performance of the FGARCH-X models does not differ significantly from that of the models using other bases. Finally, to compare the FGARCH and FGARCH-X models, the statistics overall indicate a positive sign toward the FGARCH-X models, although they are rarely significant at the 95\% level in USD/GBP and USD/JPY. These patterns are broadly retained for the inter-daily conditional volatility forecasts.
\begin{table}[H]
\centering
\caption{Results of Diebold-Mariano tests based on pairwise forecasting errors of MSFE. FGARCH, labelled FG-, and FGARCH-X, labelled X-, models are specified with four types of data-driven basis functions: truncated functional principal components (TFPCA), dynamic functional principal components (DFPCA), long-range dependent functional principal components (LFPCA), and multi-level functional principal components (MFPCA). Our first training sample of 07-January-2014 to 17-May-2017 and a forecasting sample of 18-May-2017 to 30-September-2020 are specified with a rolling forecast window dynamically updated every week, with bold values representing significance at the 5\% significance level.} \label{table-dm}
\begin{adjustbox}{max width=\linewidth}
\begin{tabular}{@{}lrrrrrr@{}}
\toprule
			& \multicolumn{3}{c}{Intraday conditional volatility forecasts} & \multicolumn{3}{l}{Inter-daily conditional volatility forecasts} \\\midrule
			Diebold-Mariano test & USD/EUR               & USD/GBP          & USD/JPY               & USD/EUR                & USD/GBP             & USD/JPY              \\\midrule
			FG-TFPCA vs FG-LFPCA & \bf{7.94} & 0.38 & 0.59 & \bf{13.52} & -1.09 & -1.31 \\
			FG-TFPCA vs FG-MFPCA & \bf{7.06} & -0.11 & \bf{4.40}  & \bf{6.97} & -0.87 & -0.91 \\
			FG-DFPCA vs FG-LFPCA & \bf{6.97} & -0.04 & 0.34  & \bf{9.50} & -1.32 & -1.25 \\
			FG-DFPCA vs FG-MFPCA & \bf{6.83} & -0.31 & \bf{3.91}   & \bf{6.00} & 1.04 & -0.91 \\
			FG-LFPCA vs FG-MFPCA & \bf{4.87} & -1.00 & 0.94  & \bf{4.96} & 1.18 & 0.12  \\
			X-TFPCA vs X-LFPCA  & -0.50 & \bf{3.12} & 0.98    & 1.05 & -0.67 & 1.34 \\
			X-TFPCA vs X-MFPCA  & 1.23 & 1.29 & 1.13 & 1.77 & -1.35 & 1.40 \\
			X-DFPCA vs X-LFPCA & 0.04 & \bf{2.82} & 0.96 & 1.23 & -1.15 & -1.28 \\
			X-DFPCA vs X-MFPCA  & 1.28 & 0.68 & 1.12  & 1.75 & -1.52 & 1.41 \\
			X-LFPCA vs X-MFPCA  & 1.35 & -0.15 & 3.19  & 1.73 & -1.66 & 0.31 \\
            FG-LFPCA vs X-LFPCA  & \bf{7.65} & 1.07 & 1.50 & \bf{2.96} & 0.35 &  1.00 \\
            FG-MFPCA vs X-MFPCA  & \bf{5.84} & 1.03 & \bf{2.50} & 0.15 & -0.64 &  1.39 \\\bottomrule
\end{tabular}
\end{adjustbox}
\end{table}

\subsection{Comparison with realised volatility-based models}

Next, to evaluate the benefits of incorporating intraday trading movements for inter-daily volatility forecasts, we also compare the proposed FGARCH-type models with nine traditional volatility models, including the GARCH(1,1), GJR-GARCH(1,1), GARCH-X, Fractionally Integrated-GARCH (FIGARCH), Heterogeneous Autoregressive (HAR) \citep{C09}, HAR-X, HAR-Autoregressive Fractionally Integrated Moving Average(ARFIMA), Realised GARCH (RGARCH) \citep{HHS2012}, and the GARCH-MIDAS \citep{ EGS2013}. Note that these comparisons only apply to inter-daily volatility forecasts.

Regarding the model selection, we specify the models with parameterisation and covariates usually used and discussed in the literature. For example, in the GARCH-X and HAR-X models, we use the inter-daily bid-ask spread as the covariate X, which is given by
\begin{equation*}
\mbox{Bid-ask Spread}_t = \frac{\mbox{ask}_t-\mbox{bid}_t}{\mbox{mid}_t},
\end{equation*}
where $\mbox{mid}_t$ is the average of the bid and ask prices on day $t$. The fractional integration order $d$ in FIGARCH(1,d,1) and HAR-ARFIMA(1,d,1) are optimally selected using the functions from {\it rugarch} and {\it arfima} in the statistical software \Rlogo . Also, in realised volatility-based models, we obtain realised volatility by summing squared intraday log returns with the squared overnight return, similar to how it was used in the main manuscript. Following \cite{EGS2013}, we use monthly realised volatility for the GARCH-MIDAS model to capture long-term variations. The realised volatility-based models are mainly implemented using functions from the packages {\it HARModel} \citep{Sjoerup19}, {\it rugarch} \citep{Galanos24} and {\it rumidas} \citep{Candila24} in statistical software \Rlogo \ \citep{Team24}. Similar to Section 5 of the main manuscript, we use the entire sample from 07-January-2014 to 30-September-20 and train each model candidate with a rolling window of 600 observations, resulting in the out-of-sample period being from 18-May-2017 to 30-September-2022. Similarly, each model is re-estimated every week to adapt to market changes. From the eight FGARCH-type models, we select the FGARCH-MFPCA model and the FGARCH-X models using the LFPCA and MFPCA bases as representative models for this comparison.

\begin{figure}[H]
\centering
\caption{Plots of realised volatility and inter-daily volatility forecasts obtained from ten model candidates, including the MFPCA-embedded FGARCH-X models, and the GARCH(1,1), GJR-GARCH(1,1), GARCH-X, FIGARCH, HAR, HAR-X, HAR-ARFIMA, RGARCH, GARCH-MIDAS models. From top to bottom, the three subplots represent the outputs for USD/EUR, USD/GBP, and USD/JPY, respectively. Our first training sample of 07-January-2014 to 17-May-2017 and a forecasting sample of 18-May-2017 to 30-September-2020 are specified with a rolling forecast window that is dynamically updated weekly.}\label{figure-models}
\includegraphics[width=14.5cm]{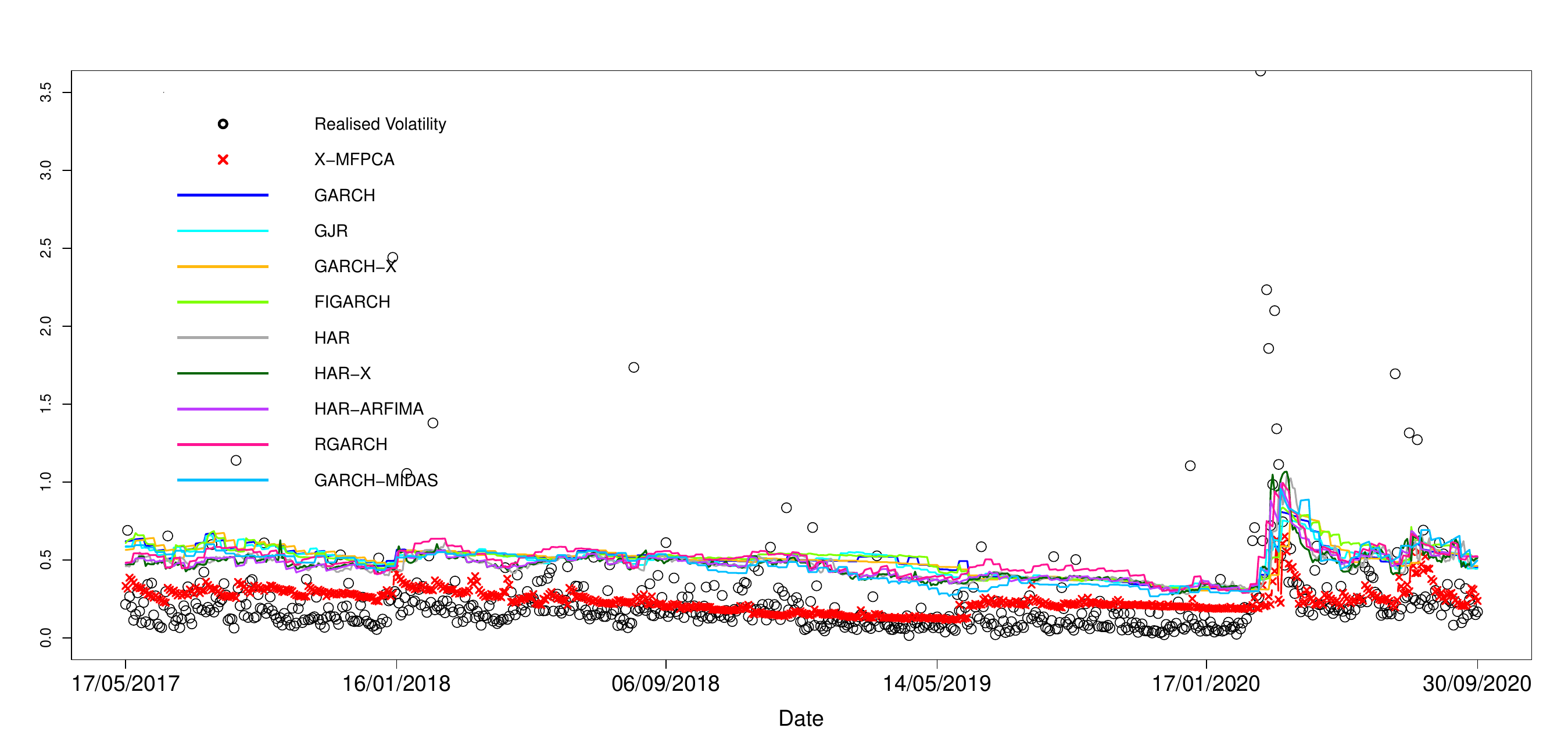}
\includegraphics[width=14.5cm]{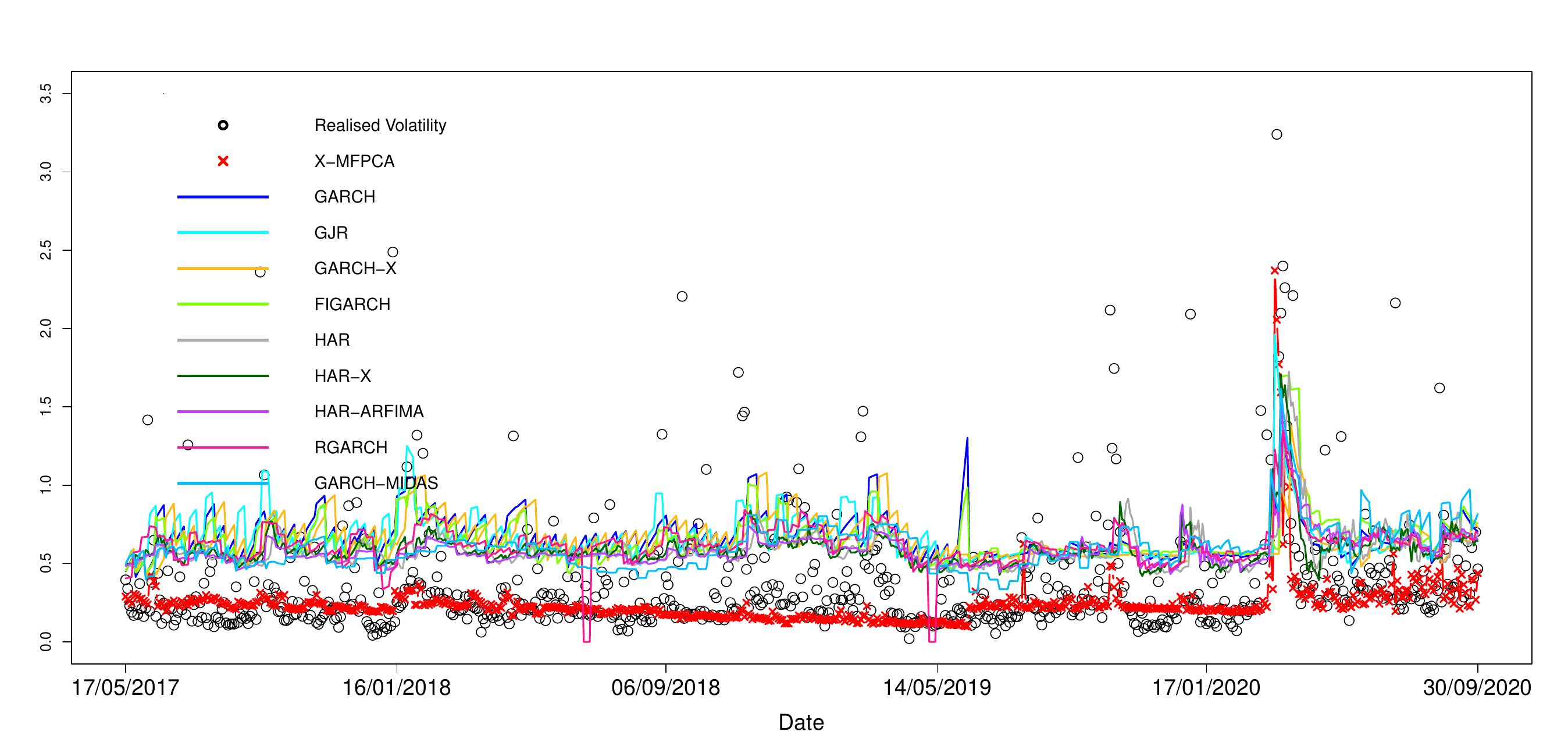}
\includegraphics[width=14.5cm]{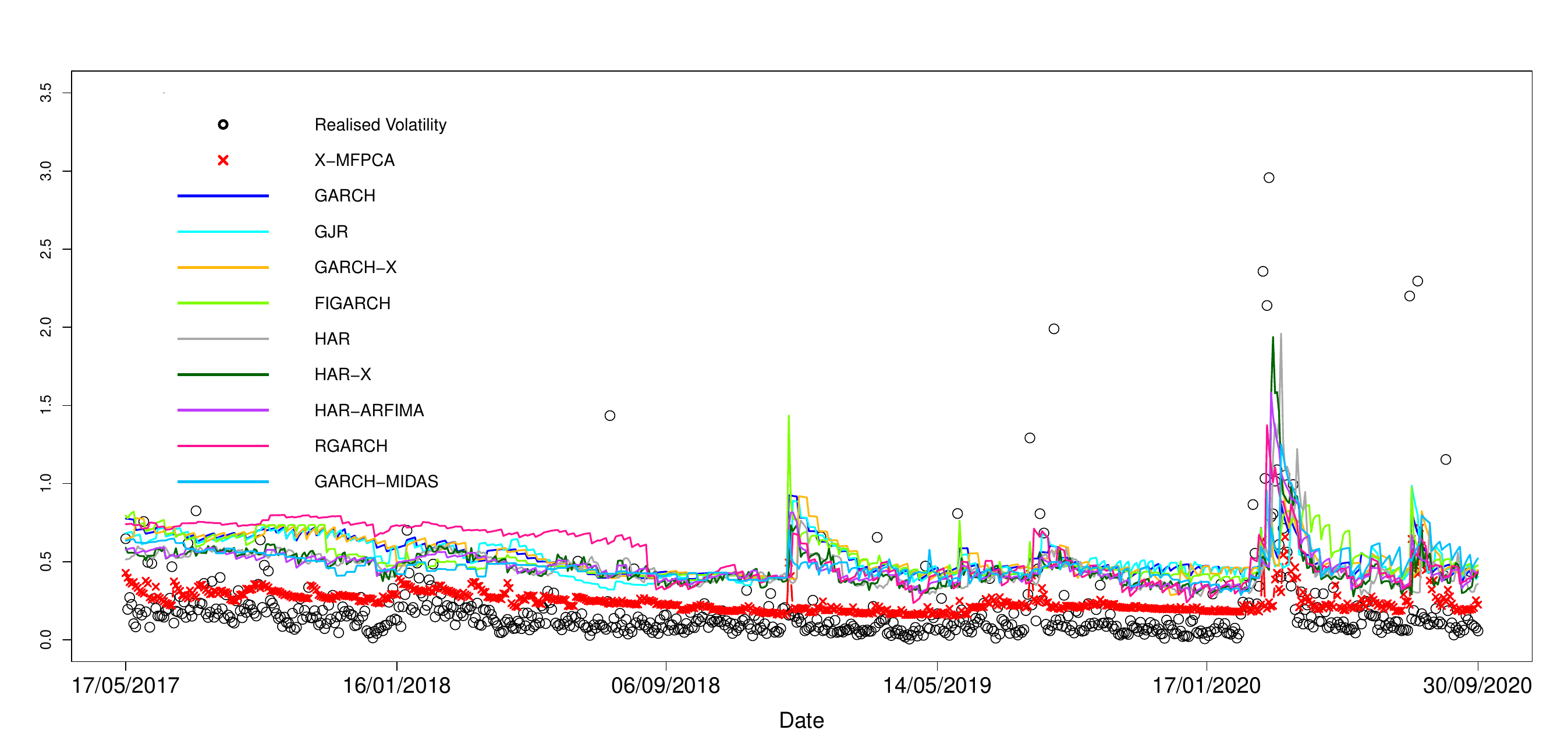}
\end{figure}

Table~\ref{table-error-models} exhibits the MSFE and QLIKE forecasting errors of eleven model candidates, which are calculated by using~\eqref{eq-forecasterror-daily} from the main context with the realised volatility estimating the true volatility. Except for the HAR-X model that produces the smallest MFSE error for USD/GBP, the overall results indicate that the FGARCH-X model outperforms other model candidates. The realised volatility-based models generally outperform scalar GARCH models, which also beat the FGARCH model. The FGARCH model shows comparable performances to the scalar GARCH models for USD/GBP and USD/JPY, but the errors are off-scale for USD/EUR. In addition, we observe that, apart from the FGARCH-X models, the HAR-X model outperforms in producing smaller forecasting errors, further supporting the predictability of bid-ask spreads to FX volatility forecasting. Comparatively, by incorporating all the intraday trading information, the FGARCH-X shows further superiority. Figure~\ref{figure-models} displays the plots of realised volatility and inter-daily volatility forecasts for nine model candidates, along with the FGARCH-X-MFPCA, across three FX pairs. The plot patterns are consistent with the findings revealed in Table~\ref{table-error-models} that the FGARCH-X volatility forecasts are closer to the realised volatility.
\begin{table}[H]
\centering
\tabcolsep 0.245in
\caption{One-step-ahead forecasting errors. Our first training sample of 07-January-2014 to 17-May-2017 and a forecasting sample of 18-May-2017 to 30-September-2020 are specified with a rolling forecast window dynamically updated every week, with bold values indicating significance at the 5\% significance level.}\label{table-error-models}
\begin{adjustbox}{max width=\linewidth}
\begin{tabular}{@{}lcccccc@{}}
\toprule
              &        \multicolumn{2}{c}{USD/EUR} & \multicolumn{2}{c}{USD/GBP} & \multicolumn{2}{c}{USD/JPY} \\\hline
               & MSFE         & QLIKE        & MSFE         & QLIKE        & MSFE         & QLIKE        \\\hline
GARCH(1,1)    & 0.1522 & -0.2634 & 0.3243 & 0.1566 & 0.8089 & -0.2257  \\
GJR-GARCH         & 0.1473 & -0.2707 & 0.3074 & 0.1508 & 0.8076 & -0.2327  \\
GARCH-X        & 0.1575 & -0.2491 & 0.3415 & 0.1729 & 0.8203 & -0.2118 \\
FIGARCH        & 0.1592 & -0.2514 & 0.3193 & 0.1478 & 0.8039 & -0.2413 \\
HAR            & 0.1318 & -0.3037 & 0.2839 & 0.1066 & 0.7749 & -0.3085 \\
HAR-X          & 0.1260 & -0.3150 & 0.2625 & 0.0926 & 0.7680 & -0.3187  \\
HAR-ARFIMA     & 0.1276 & -0.3112 & 0.2672 & 0.0944 & 0.7673 & -0.3133 \\
RGARCH         &  0.1424 & -0.2771 & 0.2841 & 0.1685 & 0.8225 & -0.2326 \\
GARCH - MIDAS  & 0.1356 & -0.3124 & 0.2795 & 0.1034 & 0.7742 & -0.2906 \\
FG-MFPCA &  0.3398  &  -0.0435  & 0.3528 & 0.1034 &  0.7900  &  -0.2572  \\
 X-LFPCA & 0.0781 & -0.4855 & \bf{0.2084} & \bf{0.0830} & 0.7228 & \bf{-0.5045} \\
 X-MFPCA & \bf{0.0740} & \bf{-0.5155} & 0.2293 & 0.1787 & \bf{0.7099} & -0.4847 \\
\bottomrule  
\end{tabular}
\end{adjustbox}
\end{table}

\begin{table}[!htb]
\centering
\tabcolsep 0.58in
\caption{Results of Diebold-Mariano tests based on pairwise forecasting errors of MSFE of inter-daily conditional volatility forecasts. Our first training sample of 07-January-2014 to 17-May-2017 and a forecasting sample of 18-May-2017 to 30-September-2020 are specified with a rolling forecast window that is dynamically updated every week, with bold values being significant at the 5\% significance level.} \label{table-dm-models}
	\begin{adjustbox}{max width=\linewidth}
	{\footnotesize	
        \begin{tabular}{@{}lrrr@{}}
\toprule 
			Diebold-Mariano test & USD/EUR               & USD/GBP          & USD/JPY                          \\\midrule
         GARCH vs GJR-GARCH & \bf{4.77} & 0.50 & 0.42 \\
	 GARCH vs GARCH-X & \bf{-2.41} & -0.64 & -1.57 \\
	 GARCH vs FIGARCH & \bf{-5.93} & 0.90 & -0.94 \\
			GARCH vs HAR & \bf{7.87} & 1.29 & 0.49 \\
			GARCH vs HAR-X & \bf{9.27} & 1.90 & \bf{3.51} \\
	 GARCH vs HAR-ARFIMA &  \bf{11.25} & 1.79 & \bf{5.29} \\
			GARCH vs RGARCH & \bf{3.93} & 1.88 & -1.21 \\
			GARCH vs MIDAS  & \bf{4.92} & 1.23 & \bf{4.49} \\
           GJR-GARCH vs GARCH-X & \bf{-3.37} & -0.90 & -0.78 \\
   GJR-GARCH vs FIGARCH & \bf{-7.31} & -0.16 & -0.72 \\
          GJR-GARCH vs HAR & \bf{6.12} & 0.46 & 0.16 \\
        GJR-GARCH vs HAR-X & \bf{8.37} & 0.92 & 1.21 \\
  GJR-GARCH vs HAR-ARFIMA & \bf{10.62} & 0.81 & 1.47 \\
            GJR-GARCH vs RGARCH  & \bf{2.42} & 0.76 & \bf{-2.75} \\
           GJR-GARCH vs MIDAS  & \bf{2.72} & 0.48 & 1.18 \\
                GARCH-X vs FIGARCH & -0.09 & 0.97 & -0.27 \\
 	GARCH-X vs HAR & \bf{6.32} & \bf{4.01} & 0.82 \\
 GARCH-X vs HAR-X & \bf{6.38} & \bf{3.69} & \bf{4.43} \\
 GARCH-X vs HAR-ARFIMA & \bf{7.05} & \bf{4.03} & \bf{6.55} \\
 GARCH-X vs RGARCH & \bf{3.45} & \bf{4.25} & -0.69 \\
 GARCH-X vs MIDAS  & \bf{5.21} & \bf{2.79} & \bf{6.22} \\
			FIGARCH vs HAR & \bf{8.95} & 0.80 & 0.85 \\
			FIGARCH vs HAR-X  & \bf{9.60} & 1.36 & \bf{3.37} \\
	 FIGARCH vs HAR-ARFIMA & \bf{11.43} & 1.26 & \bf{4.56} \\
			FIGARCH vs RGARCH & \bf{5.37} & 1.23 & -0.36 \\
	 FIGARCH vs MIDAS  & \bf{8.05} & 0.81 & \bf{3.75} \\
	 HAR vs HAR-X & \bf{2.46} & 1.31 & 0.95 \\
			HAR vs HAR-ARFIMA & \bf{2.43} & 1.04 & 1.22 \\
	 HAR vs RGARCH  & \bf{-4.74} & 0.78 & -1.03 \\
      HAR vs MIDAS & \bf{-3.15} & 0.17 & 0.94 \\
	 HAR-X vs HAR-ARFIMA   & -0.26 & -0.65 & 0.88 \\
			HAR-X vs RGARCH  & \bf{-9.49} & -0.85 & \bf{-2.42} \\
			HAR-X vs MIDAS & \bf{-4.21} & \bf{-2.21} & -0.31 \\
         HAR-ARFIMA vs RGARCH & \bf{-9.33} & -0.48 & \bf{-2.71} \\
	 HAR-ARFIMA vs MIDAS  & \bf{-4.80} & \bf{-1.93} & \bf{-2.13} \\
	  RGARCH vs MIDAS & 0.43 & -0.71 & \bf{2.50} \\
             X-MFPCA vs GARCH & \bf{-4.53} & 1.12 & -1.55 \\
		 X-MFPCA vs GJR-GARCH & \bf{-3.94} & 1.83 & -1.49 \\
			X-MFPCA vs GARCH-X & \bf{-5.51} & 0.91 & -1.71 \\
			X-MFPCA vs FIGARCH & \bf{-5.31} & 1.27 & -1.66 \\
             X-MFPCA vs HAR & -1.84 & 1.61 & -1.02 \\
			X-MFPCA vs HAR-X & -1.26 & 1.82 & -0.59 \\
	 X-MFPCA vs HAR-ARFIMA & -1.33 & 1.79 & -0.46 \\
                X-MFPCA vs RGARCH & \bf{-2.90} & 1.76 & \bf{-2.21} \\
	 X-MFPCA vs MIDAS & \bf{-2.90} & 1.60 & -0.66 \\
   \bottomrule
		\end{tabular}
        }
	\end{adjustbox}
\end{table}

Table~\ref{table-dm-models} documents the results of the Diebold-Mariano tests to demonstrate further the pairwise relative performances across forty-five pairs of models. Considering that the FGARCH-X model produces smaller forecasting errors than the FGARCH model, we perform pairwise tests with the FGARCH-X model. We calculate the statistics using the MSFE loss function. The remarkable findings can be summarised as follows. First, the FGARCH-X model generally outperforms most scalar GARCH and realised volatility-based models for USD/EUR and USD/JPY, with some exceptions for USD/GBP. This result is consistent with the results in Table~\ref{table-error-models}, indicating the benefit of taking intraday return and bid-ask spread information into account. Second, the realised volatility-based models generally show significant outperformance over the scalar GARCH-type models, which is consistent with most empirical results in the literature. Third, the exogenous covariate of bid-ask spreads is more effective in improving the forecasting accuracy of the HAR models, whereas no improvement is observed for the scalar GARCH models. Lastly, by taking into account the long-term monthly realised volatility, the GARCH-MIDAS model shows some forecasting superiority, but it barely beats the HAR-X, HAR-ARFIMA and FGARCH-X models. In a nutshell, FGARCH-type models, including the FGARCH-X model, provide competitive predictions of the inter-daily conditional volatility forecasts compared with classic time series volatility models. 

\section{A VaR corrected intraday trading strategy}\label{sec-port}

In this section, we explore a real market application in intraday risk management by taking advantage of intraday conditional volatility forecasts to uncover economic insights. The specification of the FGARCH model implies that the intraday VaR curve at quantile $\zeta$ can be calculated via,
\begin{equation}
\widehat{\text{VaR}}^\zeta_{t+1}(u) = \widehat{\sigma}_{t+1}(u)\widehat{\varepsilon}^\zeta(u),\qquad \zeta\in (0,1),
\end{equation}
where $\widehat{\sigma}_{t+1}(u)$ is the intraday volatility forecasts obtained from the models, and $\widehat{\varepsilon}^\zeta(u)$ is the $\zeta$\textsuperscript{th} unconditional quantile of the error process. \citet{RWZ20} suggest three types of error processes for $\widehat{\varepsilon}^\zeta(u)$, including a Gaussian process, generalised Pareto distributed process, and the empirical process obtained by bootstrapping the fitted error curves. Here, we estimate $\widehat{\varepsilon}^\zeta(u)$ through bootstrapping the empirical process with 10,000 replications.

An intraday VaR curve is valid if it passes the backtests on the violation process. We define the pointwise violation process $Z_t^\zeta(u)$ as
\begin{equation*}
Z_t^\zeta(u) = \mathbb{I} \left\{ \widetilde{y}_t(u)<\widehat{\text{VaR}}_t^{\zeta}(u) \right\}, 
\end{equation*}
where $\mathbb{I}\{\cdot\}$ is a binary indicator function so that the process $Z_t^\zeta(u)$ is composed of binary values, with one standing for pointwise exceeding the intraday interval. The intraday VaR curves are valid only if the process $Z_t^\zeta(u)$ is unbiased from the nominal level $\zeta$ and independent of $t$. To evaluate the validity of intraday VaR forecasts as well as to guide long and short trading strategies, we test the unbiasedness and independence hypotheses when VaR curves are computed at nominal quantiles $\zeta = 0.01$ and $0.99$ for potential intraday risk management tools.

The left part of Table~\ref{table-var-01} documents the backtest results for intraday VaR forecasts at nominal quantiles $\zeta = 0.01$. Almost all model candidates using the proposed bases produce unbiased intraday VaR curves, and their violation processes cannot reject the independence hypothesis at lags $H=1$, $5$, $10$ and $20$. The only exception occurs for the asset USD/JPY where the intraday VaR associated with the LFPCA basis function cannot pass the unbiasedness test. Overall, the results indicate valid intraday VaR forecasts, which can be used as risk management tools for intraday trading strategies. We use the intraday VaR of $\zeta=0.01$ and $\zeta=0.99$ to control for tail risks in long-and short-trading strategies. These findings generally remain consistent for $\zeta=0.99$, so we skip reporting similar results.

To further investigate the effect of extreme quantiles on intraday VaR, the right part of Table~\ref{table-var-01} displays the backtest results for $\zeta=0.0025$. The validity of our intraday VaR forecasts generally remains for USD/EUR and USD/JPY, but it deteriorates for USD/GBP. Most forecasting models cannot pass the unbiasedness backtest, although the corresponding violation processes exhibit the desirable independent property. Despite this, the findings of $\zeta = 0.0025$ are roughly consistent with those presented for $\zeta = 0.01$.
\begin{table}[H]
\centering
\tabcolsep 0.12in
	\caption{$P$-values of the unbiasedness and independence backtests for intraday VaR curves with nominal quantile $\zeta=0.01$ and $0.0025$. FGARCH and FGARCH-X models are specified with the proposed data-driven basis functions: long-range dependent functional principal components (LFPCA) and multi-level functional principal components (MFPCA). Our first training sample of 07-January-2014 to 17-May-2017 and a forecasting sample of 18-May-2017 to 30-September-2020 are specified with a rolling forecast window that is dynamically updated weekly. The lag lengths of the independence test are set at $H=1$, $5$, $10$ and $20$, with bold values being significant at the 5\% significance level.} \label{table-var-01}
	\begin{adjustbox}{max width=\linewidth}
		\begin{tabular}{@{}lccccccccc@{}}
\toprule
			&      & \multicolumn{4}{c}{ 1\%}   & \multicolumn{4}{c}{ 0.25\%} \\\cmidrule{2-10}
			&      & FG-LFPCA & FG-MFPCA & X-LFPCA & X-MFPCA & FG-LFPCA & FG-MFPCA & X-LFPCA & X-MFPCA \\\midrule
			& \multicolumn{9}{c}{Panel A: USD/EUR}                                \\\midrule
			Unbiasedness                  & $H$    &  0.11 & 0.17 & 0.08 & 0.19  & 0.34 & 0.21  &   0.35 & 0.41 \\\midrule
			\multirow{4}{*}{Independence} & $1$    & 0.25 & 0.23 &  0.32 & 0.33   & 0.30 & 0.31   &  0.21 & 0.37 \\
            & $5$   & 0.26 & 0.25 &  0.30 & 0.27  & 0.29 & 0.30  &   0.42 & 0.37   \\
			& $10$   & 0.27 & 0.25 &  0.31 & 0.27  &  0.33 & 0.29   &   0.39 & 0.36 \\
			& $20$   & 0.28 & 0.26 &  0.31 & 0.29  &  
 0.33 & 0.31 &  0.32 & 0.33 \\\midrule
			& \multicolumn{9}{c}{Panel B: USD/GBP}                                \\\midrule
			Unbiasedness                  &  $H$   & 0.45 & 0.43 &  0.35 & 0.07   &  \bf{0.01} & \bf{0.01}  & 0.08 & \bf{0.00}
  \\\midrule
			\multirow{4}{*}{Independence} & $1$   & 0.30 & 0.40 &  0.38 & 0.20   &  \bf{0.00} & 0.29   & 0.38 & 0.32  \\
			& $5$   & 0.39 & 0.35 & 0.41 & 0.20  & 0.35 & 0.36   & 0.40 & 0.36 \\
			& $10$   & 0.38 & 0.46 & 0.42 & 0.21  &  0.38 & 0.43  &  0.40 & 0.39 \\
			& $20$    & 0.43 & 0.41 &  0.43 & 0.27  & 0.37 & 0.39 &  0.39 & 0.37 \\\midrule
			& \multicolumn{9}{c}{Panel C: USD/JPY}                                \\\midrule
			Unbiasedness                 &  $H$   & \bf{0.02} & 0.13 &  0.10 & 0.18  &  0.37 & 0.39  &  0.26 & 0.50  \\\midrule
			\multirow{4}{*}{Independence} & $1$   & 0.41 & 0.35 &  0.88 & 0.42  & 0.51 & 0.74   &  1.00 & 1.00  \\
			& $5$    & 0.74 & 0.33 & 0.40 & 0.46 &   0.77 & 1.00   &  0.59 & \bf{0.00}  \\
			& $10$   & 0.38 & 0.45 &  0.34 & 0.27  &  0.41 & 0.49  &  0.08 & 0.36 \\
			& $20$    & 0.43 & 0.45 &  0.39 & 0.43  &   0.44 & 0.42  &  0.44 & 0.34\\\bottomrule
		\end{tabular}
	\end{adjustbox}
\end{table}

We now discuss an intraday asset allocation problem that benefits from constructing valid intraday VaR curves. Here, we aim to examine whether intraday VaR provides a sufficient risk management buffer to prevent the large economic losses inherent in intraday trading strategies. Thus, we introduce a simple VaR-corrected intraday trading strategy. The principle is that for each trading day, we open a position at the minimum point of the predicted OCIDR curve and close the position at the maximum point of the predicted OCIDR curve before the close of trading. When considering a long-trading strategy, the asset is bought at the predicted minimum point and sold at the maximum point, and vice versa for a short-trading strategy. The strategy is only executed when the range of $\widehat{y}_{t+1}(u)$ exceeds its mean. Otherwise, active trading is suppressed and the return is zero on that trading day.

To forecast the OCIDR curve, we use the simple historical functional mean as the OCIDR curve forecasts. Specifically, the forecast at $t+1$ is taken as the functional mean of past OCIDR curves over 102 days:
\begin{equation}\label{model-fmean}
\widehat{y}_{t+1}(u) = \frac{1}{102}\sum_{s=t-102}^{t} \widehat{y}_{s}(u).
\end{equation}
In further unreported analysis, we calculate historical functional means with lagged periods of a quarter (51 days), half-year (102 days), one year (204 days), and two years (408 days) based on four trading days a week. The results indicate improved performance when the half-year look-back period is deployed.

\begin{figure}[H]
\centering
\caption{Example of a VaR-corrected intraday long trading strategy for USD/GBP and the cumulative returns of benchmark and VaR-corrected trading strategies. The upper subplot shows that Within each trading day, we open a position at the Overnight Cumulative Intraday Return (OCIDR) curve's predicted minimum and close at its maximum if the OCIDR does not hit the intraday VaR curve. The intraday VaR is only considered when there is an open position, and the position closes immediately when a violation occurs. The upper sub-figure shows the long trading strategy, where the asset is bought at the predicted minimum and sold at the violation. The intraday time is given in the $x$-axis, with the left-hand side $y$-axis showing the intraday return and VaR and the right-hand side $y$-axis showing the functional mean. The lower subplot displays the cumulative returns of the benchmark and VaR-corrected strategies, where decisions are made based on the FGARCH and FGARCH-X models using MFPCA bases.}\label{figure-trade}
\includegraphics[width=16.5cm]{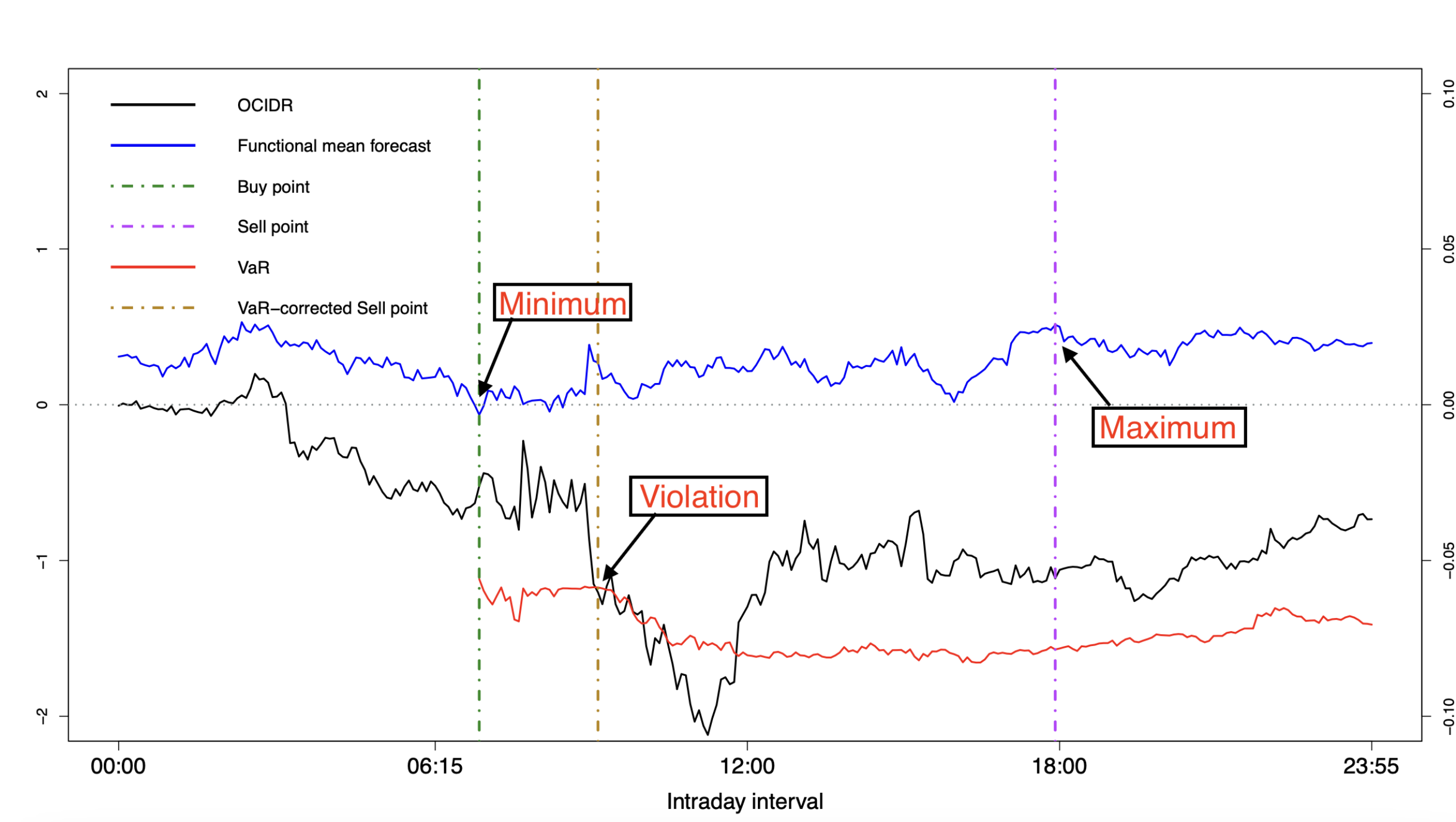}
\includegraphics[width=16.5cm]{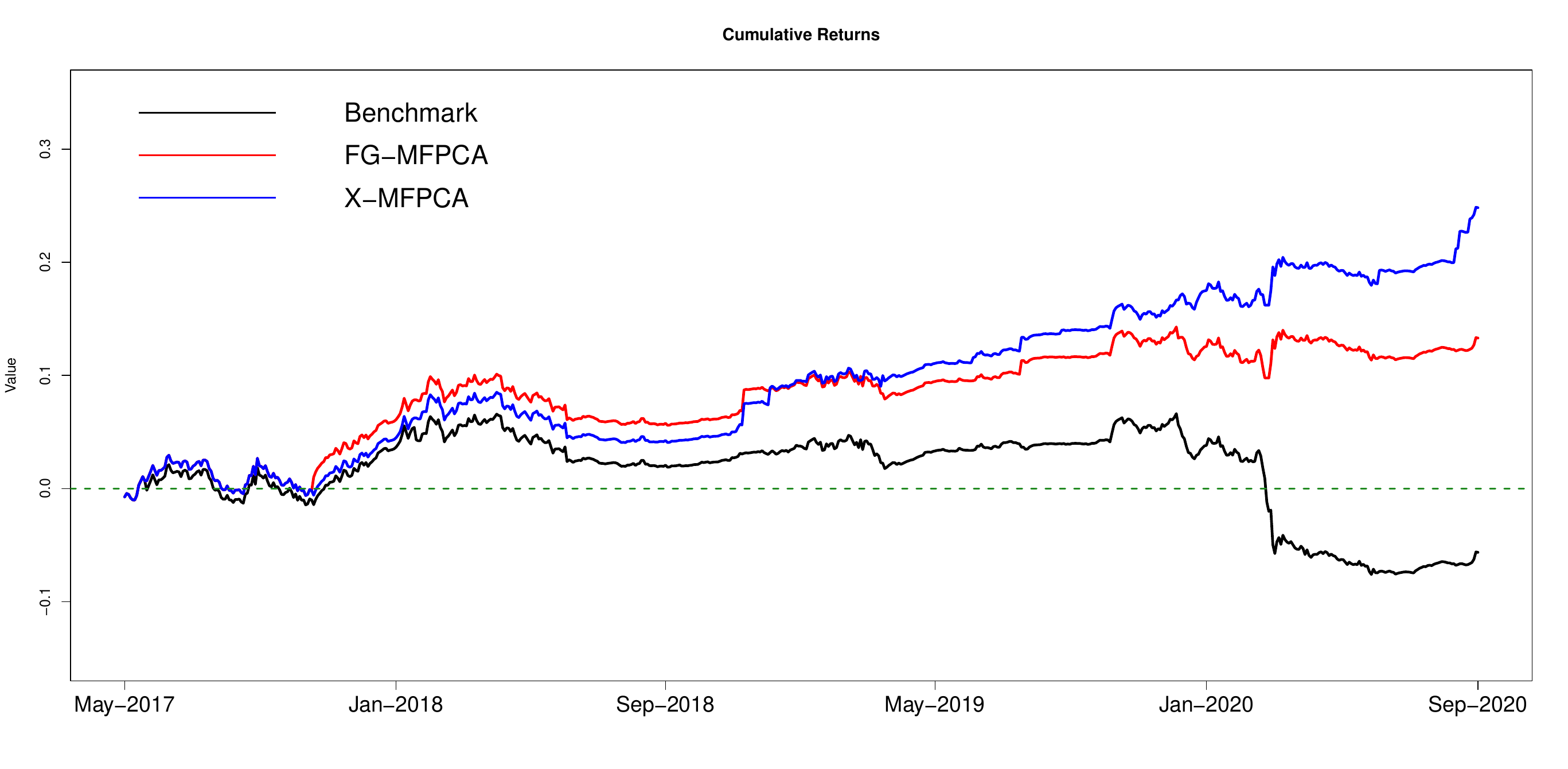}
\end{figure}

Next, we incorporate the VaR forecasts to mitigate risk during trading. Inspired by the idea of interval-valued information set, we only trade on the day that meets the condition $\left|\max{\widehat{y}_{t+1}(u)}-\min{\widehat{y}_{t+1}(u)}\right| \geq \frac{1}{J}\sum_{j=1}^J\widehat{y}_{t+1}(u_j)$. On each active trading day with open trading positions, if the OCIDR realisation hits our intraday VaR forecasts, we close the position when the violation occurs. To mimic actual trading in practice, we apply the usual FX market transaction cost of $0.0003\%$ in the simulation. The entire strategy, therefore, is as detailed in Algorithm~1 below. The upper subplot of Figure~\ref{figure-trade} shows examples of a long trading day, illustrating the scenarios that trigger VaR corrections on the benchmark trading strategy.
\begin{algorithm}\label{algo-1}
\caption{A VaR-corrected intraday trading strategy -- Steps:}\label{alg:cap}
\begin{algorithmic}[1]
   \STATE Apply the functional mean model~\eqref{model-fmean} to forecast the one-step-ahead OCIDR curve $\widehat{y}_{t+1}(u)$.
   \STATE Check the condition $|\max{\widehat{y}_{t+1}(u)}-\min{\widehat{y}_{t+1}(u)}| \geq \frac{1}{J}\sum_{j=1}^J\widehat{y}_{t+1}(u_j)$. If satisfied, proceed to the next step; otherwise, trading is suppressed on $\{t+1\}$, and the return is zero.
    \STATE Identify the minimum point on the forecast curve $\widehat{y}_{t+1}(u)$, denote it as $u^{\mbox{min}}$, which is considered the buy signal for extended trading or the sell signal for short trading on the day $\{t+1\}$.
    \STATE Identify the maximum point after $u^{\mbox{min}}$ and denote it as $u^{\mbox{max}}$, which is considered the sell signal for short trading or the buy signal for the extended trading on the day $\{t+1\}$.
    \STATE Monitor the OCIDR realisation on day $\{t+1\}$ and compare with the intraday VaR forecasts $\text{VaR}^\zeta_{t+1}(u)$. The position is closed when the OCIDR realisation hits $\text{VaR}^\zeta_{t+1}(u)$, where $\zeta=0.01$ for long trading and $\zeta=0.99$ for short trading.
    \STATE Calculate the trading strategy return using a transaction cost rate of 0.0003\%.
\end{algorithmic}
\end{algorithm}

Return curves are presented as a percentage by multiplying by 100 as shown in~\eqref{eq-ocidr}. Adopting a more practitioner-focused convention, here we consider the actual value rather than the percentage form to calculate the trading performance statistics. Table~\ref{tab:6.4} presents the results for the benchmark and FGARCH-VaR-corrected trading strategies using MFPCA bases for USD/EUR, USD/GBP, and USD/JPY. We see improvements in annual returns, Sharpe ratios, and reduced maximum drawdown values across all currencies and models. Notably, the trading strategies for USD/GBP and USD/JPY produce positive returns after intraday VaR corrections. In Panel~A of the long trading strategy, the FGARCH-X model shows superiority in most scenarios, which is consistent with the findings of smaller forecasting errors, as discussed in Section~\ref{sec-app}. The results in Panel~B of the short trading strategy show slightly mixed preferences between FGARCH and FGARCH-X, but the general pattern aligns with the long trading performance, indicating the outperformance of VaR-corrected trading strategies.
\begin{table}[H]
\centering
    \caption{VaR-corrected trading strategy performance. Within each trading day, we open a position at the Overnight Cumulative Intraday Return (OCIDR) curve's predicted minimum and close at its maximum. Panel A shows the long trading strategy, where the asset is bought at the predicted minimum and sold at the maximum, and vice versa for the short trading strategy in Panel~B. If $\widehat{y}_{i+1}(t)$ fails to exceed its mean, the return is zero for that trading day. FGARCH, labelled FG-, and FGARCH-X, labelled X-, models are specified with multi-level functional principal components (MFPCA). Our first training sample of 07-January-2014 to 17-May-2017 and a forecasting sample of 18-May-2017 to 30-September-2020 are specified with a rolling forecast window dynamically updated every three months. The strategy's annual return (AnnualRe), Sharpe Ratio (SharpeRatio), and maximum drawdown (MaxDD) are presented.} \label{tab:6.4}
	\begin{adjustbox}{max width=\linewidth}
	\begin{tabular}{@{}lccccccccc@{}}
\toprule
		& \multicolumn{3}{c}{USD/EUR}      & \multicolumn{3}{c}{USD/GBP}      & \multicolumn{3}{c}{USD/JPY}      \\\midrule
		& \multicolumn{9}{c}{Panel A: Long trading strategy}                                                  \\\midrule
		& AnnualRe & SharpeRatio & MaxDD  & AnnualRe & SharpeRatio & MaxDD  & AnnualRe & SharpeRatio & MaxDD  \\\midrule
		Benchmark &  0.0576 & 0.0149 & -0.0245   & -0.0172 & -0.0036 & -0.1341   & 0.0037 & 0.0011 & -0.0476 \\\midrule
		FG-MFPCA  & \bf{0.0838} & \bf{0.0195} & \bf{-0.0244}     & 0.0419 & 0.0089 & -0.0445   & 0.0175 & 0.0049 & \bf{-0.0426} \\ 
		X-MFPCA   & 0.0836 & \bf{0.0195} & \bf{-0.0244}  & \bf{0.0796} & \bf{0.0159} & \bf{-0.0438} & \bf{0.0256} & \bf{0.0068} & \bf{-0.0426} \\\midrule
		& \multicolumn{9}{c}{Panel B: Short trading strategy}                                                 \\\midrule
		Benchmark & 0.0207 & 0.0053 & -0.0620    & -0.0311 & -0.0061 & -0.1595   & -0.0032 & -0.0008 & -0.0667 \\\midrule
		FG-MFPCA & \bf{0.0347} & \bf{0.0092} & \bf{-0.0497}   & 0.0209 & 0.0045 & -0.0649 & \bf{0.0263} & \bf{0.0068} & -0.0395 \\
		X-MFPCA   & 0.0326 & 0.0086 & \bf{-0.0497}    & \bf{0.0226} & \bf{0.0048} & \bf{-0.0635}  & 0.0167 & 0.0043 & \bf{-0.0392} \\\bottomrule
	\end{tabular}
    \end{adjustbox}
\end{table}

To enhance our understanding, we also plot the cumulative returns of the long trading strategy for USD/GBP as an example. The lower subplot of Figure~\ref{figure-trade} shows the comparisons between the benchmark and VaR-corrected strategies. We find that the three holding strategies generally follow a similar pattern. This is because the VaR-corrected strategy, as a risk management tool, only activates when there is a violation of the 1\% intraday VaR curve, i.e., during bad days. Otherwise, the holding returns are the same as the benchmark. The VaR-corrected strategy yields higher returns because it avoids sharp declines during extreme market events. Notably, during the outbreak of COVID-19, the benchmark experienced a dramatic decline in returns, whereas the VaR-corrected strategy successfully avoided this drop. Overall, the results show that we can provide a useful intraday risk management tool for trading activities by accurately forecasting intraday volatility.
\medskip
The implementation of the methodology applied in this article is available in the online repository: \url{https://github.com/yzhao7322/FX-Multi-FGARCH}.

\section{Conclusion}\label{sec-con}

This paper seeks to enhance intraday FX market predictions by incorporating cross-asset dependence and long-range conditional heteroskedasticity features into return curve forecasting models. Our dataset includes the intraday 5-minute frequency spot price and bid and ask rates for the world's most frequently traded currencies: USD/EUR, USD/GBP, and USD/JPY over the sample period of 07-January-2014 to 30-September-2020. Using FGARCH-type models, we precisely forecast FX intraday volatilities and highlight their practical benefits through applications in intraday risk management.

The intraday evolution of FX rate movements is conveniently cast as a functional data process in acknowledgement of the continuous underlying dynamic driving its behaviour. By observing FX intraday return curves, we find that they are long-range dependent at the second moment and highly cross-correlated. We then methodologically innovate beyond the previous functional GARCH-type frameworks proposed by \cite{AHP17} and \cite{CFH+19} by incorporating the cross-dependence structure and the long-range conditional heteroskedasticity features observed in the intraday FX data. Furthermore, we also contribute by extending traditional GARCH-style models to integrate bid-ask market microstructure information using the functional GARCH-X model. Note that our modelling framework can be generally applied to other financial markets for intraday volatility forecasting to provide risk management implications.

As a result, we demonstrate the statistical superiority of our proposed models using the model confidence set and the Diebold-Mariano test. Taking into account the cross-currency dependence structure improves the accuracy of forecasting FX intraday conditional volatility. This conclusion is based on the MFPCA data-driven method that provides a superior basis for the dimension reduction process of functional GARCH-type models, especially for the FGARCH(1,1) model. The success of the MFPCA basis may be driven by increased integration among economies and currency pairs, with each country's external imbalances being a possible explanation for the observed cross-sectional variation \citep{CRS16}. Moreover, explaining the long-range dependence is beneficial in forecasting intraday and inter-daily conditional volatility. The LFPCA basis used in FGARCH-type models moderately improves forecasting accuracy. More significantly, our results provide strong evidence to suggest that incorporating the bid-ask spread information through an FGARCH-X model enhances the predictability of conditional volatility of FX returns. The proposed models also show some outperformance in terms of inter-daily volatility forecasts compared with classic realised volatility-based models.

We also demonstrate how one might economically benefit from our intraday volatility forecasting models. We calculate intraday VaR measures through intraday conditional volatility forecasts and further show how a VaR-corrected trading strategy might protect investors. Almost systematically across all currencies and forecasting models, we find superior returns, Sharpe ratios, and maximum drawdowns through the determination of the intraday VaR risk management signal, particularly with MFPCA basis-based models consistently outperforming.

For a potentially fruitful future research avenue, one might incorporate intraday options data sources into the functional GARCH-X framework, as option prices are often interpreted as the market's expectation of future realised volatility. Exploring intraday volatility modelling and forecasting in other financial assets with distinct characteristics represents an interesting direction for future research. One could also extend the FGARCH models by applying a cube root transformation to intraday return data, which would yield a more symmetric underlying distribution and potentially lead to more precise model estimation.
 


\bibliographystyle{apalike} 
\bibliography{FX_MS_abb.bib}

\end{document}